\pdfoutput=1
\documentclass[traditabstract,longauth]{aa}

\usepackage[varg]{txfonts}
\usepackage{epsfig}
\usepackage{graphicx}
\usepackage{float}
\usepackage{array}
\usepackage{longtable}
\usepackage{pdflscape}
\usepackage{url}
\usepackage[compress]{natbib}
\usepackage[colorlinks=true,
            linkcolor=blue,
            urlcolor=blue,
            citecolor=blue]{hyperref}
\setcitestyle{aysep={},yysep={;}}
\usepackage{rotating}

\begin{document}

\title{GRB 161219B / SN 2016jca: A low-redshift gamma-ray burst supernova powered by radioactive heating}

%\subtitle{ }

\author{Z.~Cano\inst{1} 
\and L.~Izzo\inst{1}
\and A.~de~Ugarte~Postigo\inst{1,2}
\and C.~C.~Th\"one\inst{1}
\and T.~Kr\"uhler\inst{3}
\and K.~E.~Heintz\inst{4,2} 
\and D.~Malesani\inst{2,5}
\and S.~Geier\inst{6,7} 
\and C.~Fuentes\inst{8,9}
%%% ABC hereafter
\and T.-W.~Chen\inst{10}
\and S.~Covino\inst{11}
\and V.~D'Elia\inst{12,13}
\and J.~P.~U.~Fynbo\inst{2}
\and P.~Goldoni\inst{14}
\and A.~Gomboc\inst{15}
\and J.~Hjorth\inst{2}
\and P.~Jakobsson\inst{4}
\and D.~A.~Kann\inst{1}
\and B.~Milvang-Jensen \inst{2}
\and G.~Pugliese\inst{16}
\and R.~S\'anchez-Ram\'irez\inst{1}
\and S.~Schulze\inst{17}
\and J.~Sollerman\inst{18}
\and N.~R.~Tanvir\inst{19}
\and K.~Wiersema\inst{19}
%\and
%
}

\offprints{ \email{zewcano@gmail.com}}

\institute{Instituto de Astrof\'isica de Andaluc\'ia (IAA-CSIC), Glorieta de la Astronom\'ia s/n, E-18008, Granada, Spain  % 1
\and Dark Cosmology Centre, Niels Bohr Institute, University of Copenhagen, Juliane Maries Vej 30, 2100 K\o benhaven \O{}, Denmark %2
\and Max-Planck-Institut f\"ur Extraterrestrische Physik, Giessenbachstrasse 1, D-85748, Garching, Germany %3
\and Centre for Astrophysics and Cosmology, Science Institute, University of Iceland, Dunhagi 5, 107 Reykjav\'ik, Iceland %4
\and DTU Space, National Space Institute, Technical University of Denmark, Elektrovej 327, 2800, Lyngby, Denmark %5
\and Gran Telescopio Canarias (GRANTECAN), Cuesta de San Jos\'e s/n, E-38712, Bre\~na Baja, La Palma, Spain %6
\and Instituto de Astrof\'isica de Canarias, Vía L\'actea s/n, E-38200, La Laguna, Tenerife, Spain %7
\and Departamento de Astronom\'ia, Universidad de Chile, Camino del Observatorio 1515, Las Condes, Santiago, Chile %8
\and Millennium Institute of Astrophysics, Chile %9
\and Max-Planck-Institut f{\"u}r Extraterrestrische Physik, Giessenbachstra\ss e 1, 85748, Garching, Germany %10
\and INAF / Osservatorio Astronomico di Brera, via Bianchi 46, 23807, Merate (LC), Italy %11
\and INAF-Osservatorio Astronomico di Roma, Via Frascati 33, I-00040 Monteporzio Catone, Italy % 12
\and ASI-Science Data Centre, Via del Politecnico snc, I-00133 Rome, Italy % 13
\and APC, Astroparticule et Cosmologie, Universit\'e Paris Diderot, CNRS/IN2P3, CEA/Irfu, Observatoire de Paris, Sorbonne Paris Cit\'e, 10, Rue Alice Domon et L\'eonie Duquet, 75205, Paris Cedex 13, France %14
\and School of Science, University of Nova Gorica, Vipavska 11c, 5270 Ajdov\v s\v cina, Slovenia %15
\and Anton Pannekoek Institute for Astronomy, University of Amsterdam, Postbus 94249, NL-1090 GE Amsterdam, The Netherlands  %16
\and Department of Particle Physics and Astrophysics, Weizmann Institute of Science, Rehovot 76100, Israel %17
\and The Oskar Klein Centre, Department of Astronomy, AlbaNova, SE-106 91 Stockholm , Sweden  %18
\and Department of Physics and Astronomy, University of Leicester, University Road, Leicester, LE1 7RH, UK %19
}

\date{Received xx xxx 2017 / Accepted xx xxx 2017}

\abstract {Since the first discovery of a broad-lined type Ic supernova (SN) with a long-duration gamma-ray burst (GRB) in 1998, fewer than fifty gamma-ray burst supernovae (GRB-SNe) have been discovered.  The intermediate-luminosity \emph{Swift} GRB~161219B and its associated supernova SN~2016jca, which occurred at a redshift of $z=0.1475$, represents only the seventh GRB-SN to have been discovered within 1~Gpc, and hence provides an excellent opportunity to investigate the observational and physical properties of these very elusive and rare type of SN.  As such, we present optical to near-infrared photometry and optical spectroscopy of GRB~161219B and SN~2016jca, spanning the first three months since its discovery.  GRB~161219B exploded in the disk of an edge-on spiral galaxy at a projected distance of 3.4~kpc from the galactic centre.  GRB~161219B itself is an outlier in the $E_{\rm p,i}-E_{\rm \gamma, iso}$ plane, while SN~2016jca had a rest-frame, peak absolute $V$-band magnitude of $M_{V} = -19.0$, which it reached after $12.5$ rest-frame days.  We find that the bolometric properties of SN~2016jca are inconsistent with being powered solely by a magnetar central engine, as proposed by other authors, and demonstrate that it was likely powered exclusively by energy deposited by the radioactive decay of nickel and cobalt into their daughter products, which were nucleosynthesized when its progenitor underwent core collapse.  We find that 0.22~M~$_{\odot}$ of nickel is required to reproduce the peak luminosity of SN~2016jca, and we constrain an ejecta mass of $5.8$~M~$_{\odot}$ and a kinetic energy of $\approx5\times10^{52}$~erg.  Finally, we report on a chromatic, pre-maximum bump in the $g$-band light curve, and discuss its possible origin.}
%SN~2016jca had a peak bolometric luminosity of $L_{\rm p}=4.6\times10^{42}$~erg~s$^{-1}$, which it reached after $10.7$~days, and the photospheric velocity at this time, as inferred from the blueshift of the Fe \textsc{ii} $\lambda$5169 transition, was $29,700\pm 1,500$~km~s$^{-1}$.  
%
%The nickel and ejecta masses are commensurate with those of other GRB-SNe, though its kinetic energy is larger, owing to its larger peak photospheric velocity relative to a typical GRB-SN.  
\keywords{TBC}
%\titlerunning{The radioactively powered SN 2016jca }
%\authorrunning{Cano et al. } 
\maketitle

\begin{figure*}
   \centering % lower-left-x (llx), lower-left-y (lly), upper-right-x (urx), and upper-right-y (ury). The width of the picture is urx-llx and the height is ury-lly.
   \includegraphics[width=\hsize ]{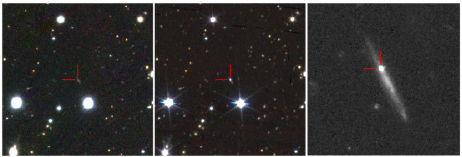}
     \caption{\textit{Left}: RGB composite Pan-STARRS1 pre-imaging of the host galaxy. \textit{Centre}: a GTC image (RGB) showing the supernova starting to emerge on the 27$^{\rm th}$ of December, 2016. \textit{Right}: \emph{HST} WFC/UVIS $F200LP$ image of the host galaxy and optical transient from the January 16$^{\rm th}$, 2017.}
         \label{FigFC}
\end{figure*}

\section{Introduction}
The connection between long-duration gamma-ray bursts (GRBs) and broad-lined type Ic supernovae (SNe IcBL) is now firmly established: i.e. the ``GRB-SN connection'' (e.g. \citealt{WoosleyBloom06,Cano2016}).  GRB-SNe are intrinsically rare; in the two decades since the association between GRB~980425 and SN~1998bw \citep{Galama98,Patat01}, fewer than 50 events have been detected \citep{Cano2016} to varying degrees of confidence \citep{HjorthBloom12}, over a wide range of redshifts ($z = 0.0087$ for GRB~980425, \citealt{LHW14}; to $z = 1.0585$ for GRB~000911, \citealt{Price02}).  Indeed only six confirmed GRB-SNe have been detected within 1~Gpc ($z \le 0.2$), and only one of these was a high-luminosity GRB\footnote{Definitions for the isotropic-equivalent luminosity in $\gamma$-rays, $L_{\gamma,\rm iso}$, e.g. \citet{Bromberg11,Hjorth13,Cano2016}, are: Low-luminosity GRB: $L_{\rm \gamma,iso} < 10^{48.5}$~erg~s$^{-1}$; Intermediate-luminosity GRB: $10^{48.5} \leq L_{\rm \gamma,iso} \leq 10^{49.5}$~erg~s$^{-1}$; and high-luminosity GRB: $L_{\rm \gamma,iso} > 10^{49.5}$~erg~s$^{-1}$.}: GRB~030329/SN~2003dh \citep{Hjorth03,Stanek03,Matheson03}; while the remainder were an intermediate-luminosity GRB: GRB~130702A/SN~2013dx \citep{Delia15,Toy16}; and low-luminosity GRBs: GRB~980425/SN~1998bw \citep{Galama98,Patat01};  GRB~031203/SN~2003lw \citep{Malesani04}; GRB~060218/SN~2006aj, \citep{Pian06,Mazzali06}; and GRB~100316D/SN~2010bh \citep{Cano2011b,Bufano12,Olivares12}.

Despite their rarity, many aspects regarding their physical properties have been established.  Their explosion mechanism is thought to be driven by the compact object that forms at the time of core-collapse: either a stellar-mass black hole (BH; \citealt{Woosley93,MacWoos1999}) or a rapidly rotating neutron-star with an exceptionally large magnetic field (a magnetar; \citealt{Usov92,ThompsonDuncan93,Bucciantini07}).  It is thought that these "central engines" lead to the explosion of the star, rather than the conventional neutrino-driven explosion mechanism (e.g. \citealt{Sukhbold16}).  From modelling observations of GRB-SNe, it has been demonstrated that their kinetic energies cluster around a value of $2-2.5 \times 10^{52}$~erg \citep{Mazzali14}, which may indicate the formation of a magnetar central engine.  However, if magnetars are formed in the majority of GRB-SNe, they are unlikely to be the dominant source of energy that powers their luminosity; instead they are very likely powered by radioactive heating \citep{CJM16}.  Next, GRB-SNe have a luminosity$-$stretch/decline relationship \citep{Cano14,LiHjorth2014,CJG14} analogous to that of type Ia SNe \citep{Phillips93}, which implies their use as cosmological probes \citep{LHW14,CJG14}.

While many aspects of the GRB-SN connection have been determined, over the years even more complex GRB-SN phenomenology has been observed.  The connection between ultra-long duration \citep{Levan14} GRB~111209A and its associated SN~2011kl \citep{Greiner15,Kann16} is one such example.  SN~2011kl is peculiar in many ways, including being exceptionally luminous (the most luminous GRB-SN to date), and whose peak optical spectrum was quite flat and featureless, where the undulations typical of GRB-SNe were conspicuously absent.  Instead, the shape of the spectrum more resembled that of superluminous supernovae (SLSNe; \citealt{Quimby11,GalYam12}), thus suggesting a link between these two types of luminous SNe, perhaps in terms of their explosion mechanism and/or processes powering the SNe themselves.  An interpretation of both the spectrum and the bolometric light curve (LC) of SN~2011kl led to the conclusion that the GRB and SN were both driven by a magnetar central engine, and radioactive heating played a minor, perhaps even negligible role in powering its luminosity \citep{Greiner15,Metzger15,Bersten16,CJM16,GompFruch17}.  A magnetar central engine has been inferred for several type Ic SLSNe \citep{Chatzopoulos11,Inserra13,Nicholl13,Chen15}, suggesting at least one common theme between these extreme stellar explosions.  As the luminosity of a magnetar-powered SN is directly related to how long the central engine is active, where central engines with longer durations give rise to brighter SNe, the key difference between the magnetar's properties in a SLSNe relative to SN~2011kl is the spin-down timescale, which is of order several weeks to months for SLSNe, but was less than a week for SN~2011kl \citep{CJM16}.  This naturally explains why SN~2011kl, though more luminous than all other GRB-SNe, was not as luminous as SLSNe-I.

In this paper we take a close look at GRB~161219B and its associated SN~2016jca.  GRB~161219B is only the seventh GRB to be detected at $z<0.2$, and the closest long-duration GRB to be detected by \emph{Swift} since GRB~100316D \citep{Starling11}.  As such, this event represents a rare opportunity to determine the detailed properties of a nearby GRB-SN.  GRB~161219B was detected by the Burst Alert Telescope (BAT) aboard \emph{Swift} \citep{Gehrels04} at 18:48:39 UT on the 19$^{\rm th}$ of December, 2016 \citep{Dai16}.  The BAT light curve showed a single-peaked structure with a duration of $\sim 10$~seconds.  The enhanced X-Ray Telescope (XRT) location was found to be RA, Dec (J2000) 06$^{\rm h}$06$^{\rm m}$51.36$^{\rm s}$ $-$26$^{\rm d}$47$^{\prime}$29.9$^{\prime\prime}$ \citep{Beardmore16}.  The GRB was also detected by Konus-\emph{Wind} \citep{Frederiks16}, who measured a duration of roughly 10 seconds, and by \emph{POLAR} \citep{Xiao16}, who measured a duration of $T_{\rm 90} = 4.0 \pm 0.5$~s.  The optical and near-infrared (NIR) afterglow (AG) was detected by many ground-based facilities \citep{Buckley16,Fong16,Fujiwara16,Guidorzi16,Kruehler16,Marshall16,Martin16,Mazaeva16}.  The AG was also detected at sub-mm \citep{Laskar16}  and radio \citep{Alexander16,Nayana16} frequencies. The redshift was determined from an optical spectrum of the AG to be $z = 0.1475$ \citep{Tanvir16,deUgarte16}.  Spectroscopic identification of its associated SN~2016jca was first given by \citet{deUgarte16}, which was later verified by a photometric SN-``bump'' by \citet{Volnova17} and additional spectroscopy by \citep{Chen17}. Finally, an in-depth investigation of the SN is presented in \citet{Ashall17}, who concluded that a magnetar central engine likely powered SN~2016jca, which is at odds to our results here, where we demonstrate that the SN was likely powered in part, or perhaps exclusively, by energy deposited by the radioactive decay of nickel and cobalt into their daughter products.

%Konus-wind determined an energy release in $\gamma$-rays in the 20-1000~keV energy range of $E_{\gamma, \rm iso} = 1.6 \times 10^{50}$~erg\footnote{for a $\Lambda$-CDM cosmological model with $H_0 = 70$ km~s$^{-1}$~Mpc$^{-1}$, $\Omega_M = 0.27$, and $\Omega_{\Lambda} = 0.73$}. 

%In this paper, we present optical spectroscopy and optical/NIR photometry of GRB~161219B and SN~2016jca. 

All data presented here have been corrected for foreground extinction using the dust maps of \citet{Schlegel98} as revised by \citet{SchFink11}.  We have assumed a generic $\Lambda$-CDM cosmological model with $H_0 = 70$ km~s$^{-1}$~Mpc$^{-1}$, $\Omega_M = 0.27$, and $\Omega_{\Lambda} = 0.73$.  The respective forward-shock afterglow (AG) decay and energy spectral indices $\alpha$ and $\beta$ are defined by $f_{\nu} \propto (t - t_{0})^{-\alpha}\nu^{-\beta}$, where $t_{0}$ is the time at which the GRB triggered \emph{Swift}-BAT, and $\nu$ is the observed frequency. 

The paper is organised as: In Section \ref{sec:datared} we present our optical and NIR photometry, including our data-reduction \& calibration procedures.  In Section \ref{sec:highenergy} we discuss the high-energy ($\gamma$- and X-ray) properties of GRB~161219B.  We present our investigation of the optical/NIR properties of the AG (Section \ref{sec:AG}) and accompanying SN (Section \ref{sec:SN_optical}), which follows our analysis of the time-resolved X-ray to NIR spectral energy distribution (SED) of GRB~161219B (Section \ref{sec:SED}).  In Section \ref{sec:bolo} we model a \emph{quasi}-bolometric LC constructed from our de-reddened (foreground and host), host- and AG-subtracted observations to determine whether radioactive-heating or a magnetar is responsible for powering its luminosity.  In Section \ref{sec:gbandLC} we discuss the intriguing presence of a chromatic pre-maximum bump present only in the $g$-band LC of SN~2016jca.  The observational and physical properties of its host galaxy are presented in Section \ref{sec:host}, and finally a discussion and our conclusions are given in Section \ref{sec:conclusions}.

\section{Data Reduction, Photometry \& Spectroscopy}
\label{sec:datared}

\subsection{Photometry}

We obtained optical and NIR observations with three ground-based telescopes: $g^{\prime}r^{\prime}i^{\prime}z^{\prime}JHK_s$ imaging with the Gamma-Ray burst Optical/Near-infrared Detector (GROND; \citealt{Greiner08}) mounted on the MPG 2.2~m telescope on La Silla, Chile; $grizJH$ imaging with the 2.5~m Nordic Optical Telescope (NOT), and $grizJHK$ imaging with the 10.4~m Gran Telescopio Canarias (GTC) telescope (OSIRIS and CIRCE), both located on La Palma, Spain.  Image reduction of the GTC, NOT and GROND \citep{Kruehler08} data was performed using standard techniques in \texttt{IRAF}\footnote{\texttt{IRAF} is distributed by the National Optical Astronomy Observatory, which is operated by the Association of Universities for Research in Astronomy, Inc., under cooperative agreement with the National Science Foundation.}, while those obtained with CIRCE \citep{Garner14} were reduced using custom codes written in IDL.   

The optical $griz$ images were calibrated to the Panoramic Survey Telescope and Rapid Response System (Pan-STARRS1; \citealt{Chambers16,Flewelling16}).  Twenty-three Pan-STARRS1 reference stars in the GRB field-of-view were chosen, and zeropoints were calculated between the instrumental magnitudes and catalog values.  The NIR images ($JHK$) were calibrated to the 2MASS catalog \citep{Kleinmann94} also using a zeropoint calculation.  Due to differences between the GROND and Pan-STARRS1 filters, an additional zeropoint correction was applied to the $g$- and $z$-band GROND images following the prescription in \citet{Finkbeiner16} (see their table 2 and equ. 1).

Inspection of the images reveals that GRB~161219B occurred in an apparent edge-on spiral galaxy (Section \ref{sec:host}; see Fig. \ref{FigFC}), meaning that the photometry of the optical transient (OT) is polluted by flux from the underlying host.  In order to provide a consistent analysis of the photometric evolution of the OT, all aperture photometry was performed using an aperture with a $2\farcs2$ radius centered on the position of the OT.

The position of the OT was determined via two methods: (1) for the first night of imaging obtained on each telescope (i.e. when the OT was brightest), we used the \texttt{IRAF} routine \texttt{centroid} within the \texttt{DIGIPHOT/DAOPHOT} package and compared the determined centroid position among the different telescopes; next (2) we used the Pan-STARRS1 images in $griz$ as templates for the image-subtraction technique, which was performed using an adaptation of the original \texttt{ISIS} program \citep{Alard98,Alard2000} that was developed for Hubble Space Telescope (\emph{HST}) SN surveys by \citet{Strolger04}, and employed in other GRB-SN studies by our group \citep{Cano2011a,Cano+14,Cano15}.  We performed image subtraction on the same epoch as method (1), and used \texttt{centroid} to find the position of the OT in the subtracted image.  We then took the average of all values, finding a position of GRB~161219B of RA, Dec (J2000) 06$^{\rm h}$06$^{\rm m}$51.412$^{\rm s}$(29) $-$26$^{\rm d}$47$^{\prime}$29.49$^{\prime\prime}$(15).

Our monitoring campaign of GRB~161219B / SN~2016jca finished at the end of February 2017 when it was no longer visible from La Palma.  As such, we were not able to obtain late-time images of the host galaxy in each filter obtained by each telescope/detector to use as templates for image subtraction.  Instead, we have quantified the amount of host flux present in the images by performing aperture photometry on the Pan-STARRS1 images using an aperture of $2\farcs2$ centered on the position quoted above.  In this sense the data presented in Fig. \ref{FigLC_optical} (right panel) are host subtracted via the flux-subtraction technique \citep{Cano14,Cano2016,Volnova2016b}.

\begin{table}
%\scriptsize
\centering
\setlength{\tabcolsep}{4.0pt}
\setlength{\extrarowheight}{3pt}
\caption{GRB 161219B / SN 2016jca: vital statistics}
\label{table:GRB_vitals}
\begin{tabular}{rl}
\hline			
GRB 161219B / SN 2016jca	&	Ref.	\\
\hline			
RA(J2000) = 06$^{h}$06$^{m}$51.412$^{s}$ 	&	this work	\\
Dec(J2000) = -26$^{d}$47$^{'}$29.49$^{"}$	&	this work	\\
$z=0.1475$	&	\citet{Tanvir16}	\\
$d_{\rm L}^{*} =  700$ Mpc	&	this work	\\
$\mu^{*} = 39.22$ mag	&	this work	\\
$E(B-V)_{\rm fore} = 0.0281 \pm 0.0002$ mag	&	\citet{SchFink11}	\\
$E(B-V)_{\rm host} = 0.017 \pm 0.012$ mag	&	this work	\\
$t_{90} =6.9$ s	& this work	\\
$E_{\rm \gamma, iso, rest} = (8.50_{-3.75}^{+8.46}) \times 10^{49}$ erg	&	this work	\\
$E_{\rm \gamma, p,rest} = 105.9_{-33.2}^{+78.2}$ keV	&	this work	\\
$L_{\rm \gamma} = (1.41_{-0.62}^{+1.41}) \times 10^{49}$ erg~s$^{-1}$ & this work \\
$v_{\rm ph, peak} = 29\,700\pm 1500$ km s$^{-1}$	&	this work, based on Fe \textsc{ii} $\lambda$5169	\\
$M_{\rm Ni} = 0.22 \pm 0.08$ M$_{\odot}$	&	this work	\\
$M_{\rm ej} = 5.8 \pm 0.3$ M$_{\odot}$	&	this work	\\
$E_{\rm K} = (5.1 \pm 0.8) \times 10^{52}$ erg	&	this work	\\
\hline					
\end{tabular}
%\medskip
\begin{flushleft}
$^{*}$ Calculated using $H_{0} = 70$ km s$^{-1}$ Mpc$^{-1}$, $\Omega_{\rm M} = 0.3$, $\Omega_{\rm \Lambda} = 0.7$. \\ 
\end{flushleft}
\end{table}

\subsection{Spectroscopy}

We obtained eight epochs of spectroscopy\footnote{All spectra presented in this paper are publically available at \url{http://grbspec.iaa.es/} \citep{deUgartegrbspec}.} of GRB~161219B and its accompanying SN~2016jca with the GTC-OSIRIS, using grisms R1000B and R1000R.  We obtained an additional spectrum of the AG-dominated OT with the X-Shooter (XS) instrument \citep{Vernet2011} mounted on Unit Telescope 2 (UT2, Kueyen) of the Very Large Telescope (VLT) at the Paranal Observatory.  We also present an optical spectrum obtained by the Public ESO Spectroscopic Survey of Transient Objects (PESSTO; \citealt{Smartt15}) that used the EFOSC2 instrument mounted on ESO's 3.58~m New Technology Telescope (NTT), obtained on 04 January, 2017 \citep{Chen17}.  The GTC and NTT spectra were reduced using standard techniques with \texttt{IRAF}-based scripts, while the XS spectra were reduced using \texttt{IRAF} and \texttt{IDL} routines.  Our spectroscopic observation log is found in Table \ref{table:spectra_obs_log}, and the spectroscopic time-series is presented in Fig. \ref{FigSpectra}.

\begin{figure}
   \centering% trim={<left> <lower> <right> <upper>}%, 
   \includegraphics[width=\columnwidth, trim={0 0 0 0}]{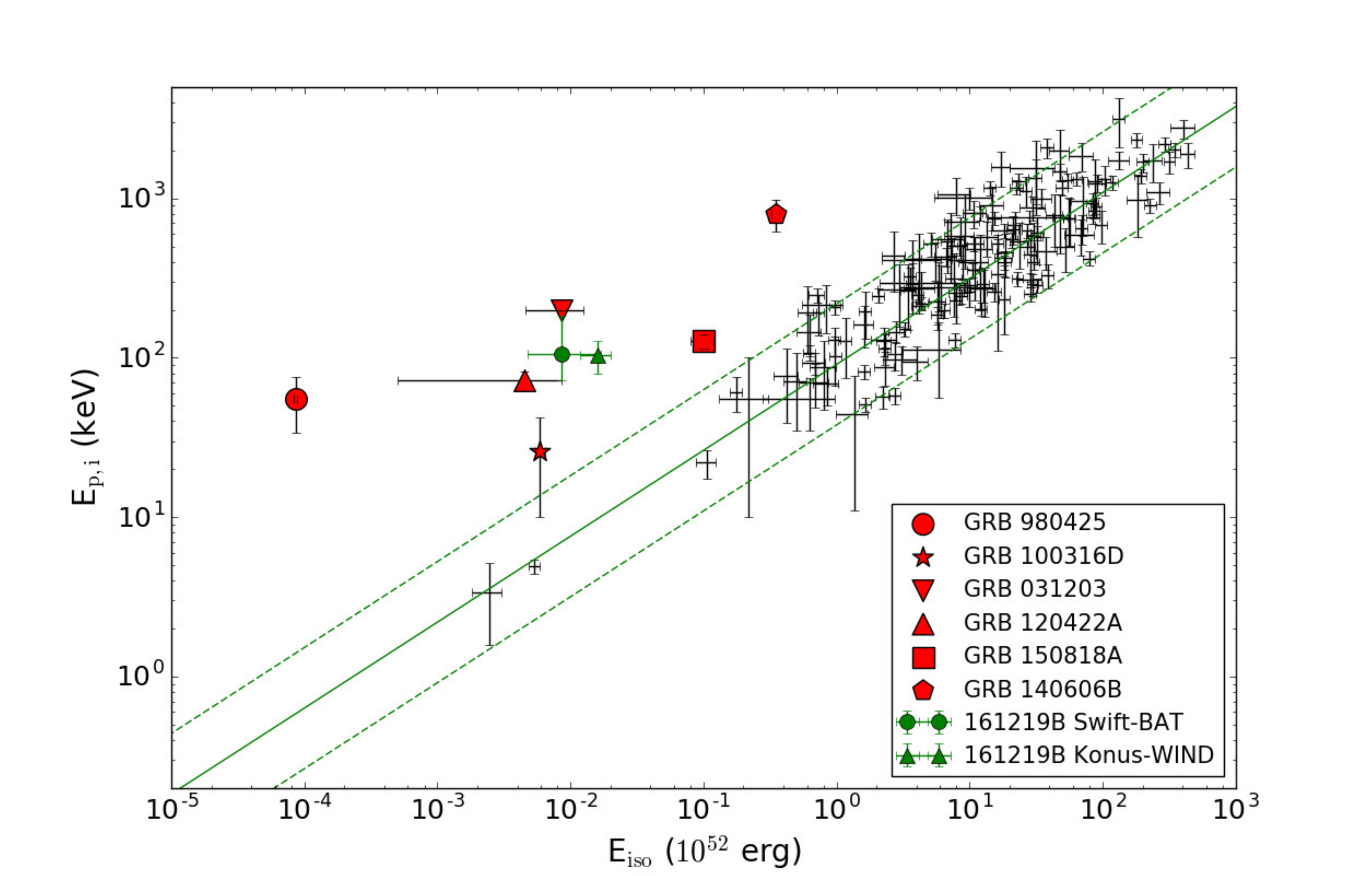}
      \caption{The position of the intermediate-luminosity GRB~161219B in the $E_{\rm p,i}-E_{\rm iso}$ (Amati) plane.  Shown for comparison is the GRB sample presented in \citet{Amati02} and \citep{Cano15}, as well as other outliers of the Amati relation, including low-luminosity GRBs 980425, 031203 \& 100316D, intermediate-luminosity GRB~150818A and high-luminosity GRBs 120422A \& 140606B.}
    \label{FigAmati}
\end{figure}

\section{The High-Energy Emission}
\label{sec:highenergy}

\subsection{Gamma-rays}

GRB~161219B was observed in $\gamma$-rays by \emph{Swift}-BAT, Konus-\emph{Wind} and by the \emph{POLAR} GRB polarimeter. We reduced the \emph{Swift}-BAT data using the standard pipeline \texttt{batgrbproduct}, and then analyzed the spectrum, integrated over the $T_{90} = 6.9$~s duration, with XSPEC \citep{Arnaud96}. The best-fitting model to the data is a single power-law (SPL) function with an exponential cutoff at the observed energy $E_{0} = 92.3_{-29.0}^{+68.2}$ keV, and a power-law photon index of $\Gamma_{\gamma} = -1.40_{-0.24}^{+0.23}$, which are in agreement with similar analysis \citep{Palmer16}. These values correspond to an intrinsic peak energy of $E_{\rm p,i} = 105.9_{-33.2}^{+78.2}$~keV and a total isotropic energy emitted in the range ($1-10000$~keV) of $E_{\rm \gamma,iso} = 8.50_{-3.75}^{+8.46} \times 10^{49}$~erg. 

These quantities indicate that GRB~161219B is an outlier in the $E_{\rm p,i}-E_{\rm \gamma, iso}$ plane (i.e. the Amati relation; \citealt{Amati02}).  In Fig. \ref{FigAmati} we have  plotted for comparison low-luminosity GRBs \citep{Cano2016,Martone17}, including GRB~980425 \citep{Galama98}, GRB~031203 \citep{Malesani04} and GRB~100316D \citep{Starling11}; intermediate-luminosity GRB~150818A \citep{Palmer15,Golenetskii15}, and high-luminosity GRBs 120422A \citep{Schulze14} and 140606B \citep{Cano15}.  We also fit the Konus-\emph{Wind} data using an identical method, and again found that GRB~161219B is an outlier in the Amati relation.  

In terms of its $\gamma$-ray luminosity, where $L_{\gamma, \rm iso} = E_{\gamma, \rm iso}~(1+z)~t_{90}^{-1}$, we find $L_{\rm, \gamma} = (1.41_{-0.62}^{+1.41}) \times 10^{49}$ erg~s$^{-1}$, and log$_{10}(L_{\rm, \gamma}) = 49.15_{-0.25}^{+0.30}$.  Using the definitions given in the introduction, GRB~161219B is an intermediate-luminosity GRB.  Other examples of intermediate-luminosity GRBs include GRB~120714B \citep{Cummings12,Klose12}, GRB~130702A \citep{Delia15,Toy16}, and GRB~150818A \citep{Palmer15,Golenetskii15,deUgarte15}.

\subsection{X-rays}

\begin{figure}
   %\centering
   \includegraphics[width=\columnwidth]{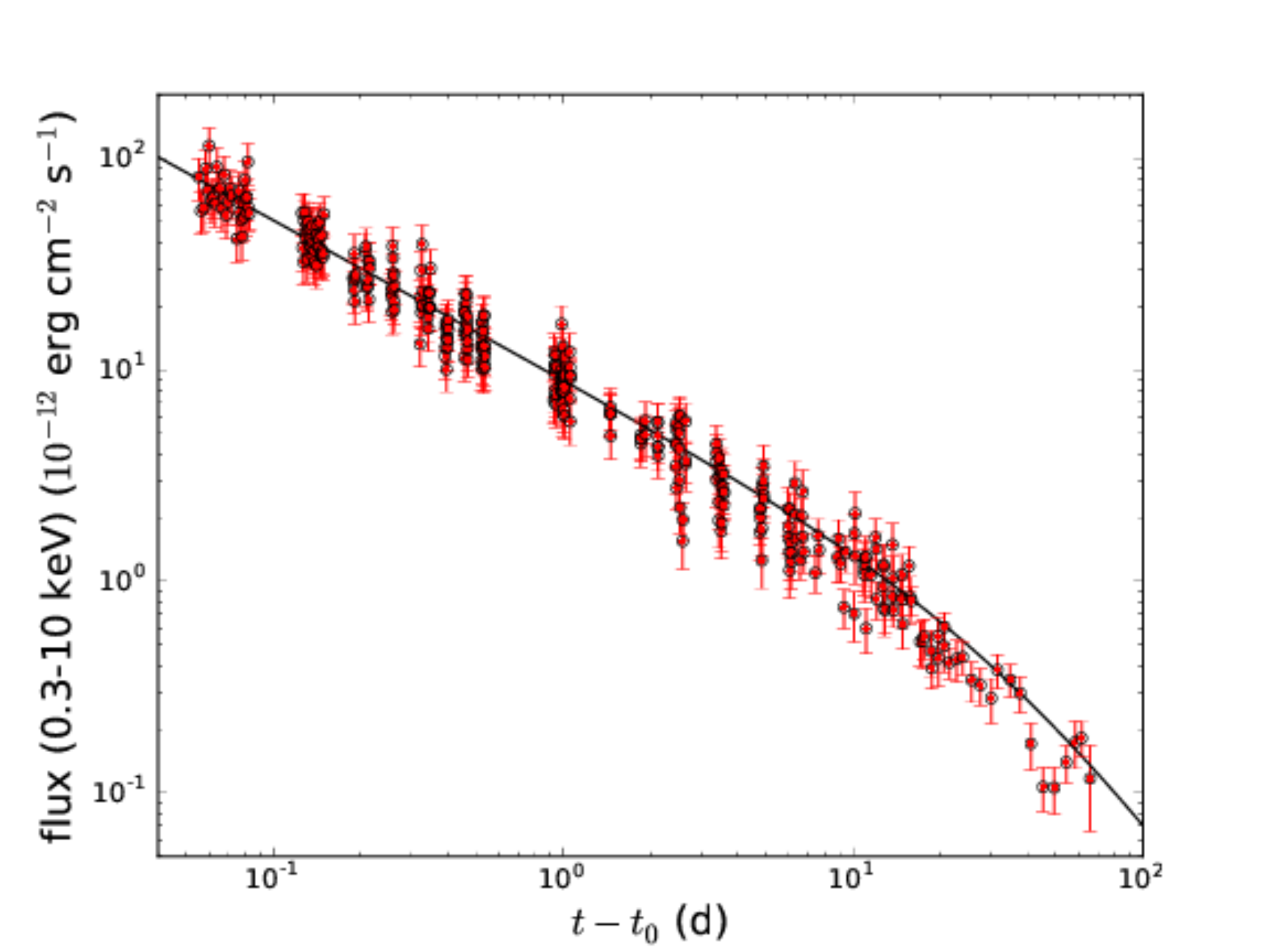}
      \caption{The \emph{Swift}-XRT LC of GRB~161219B / SN~2016jca (orbit 1 was omitted $-$ see the main text for more details).  A broken power-law was fit to the data, with the best-fitting parameters being: $\alpha_{1} = 0.79\pm0.02$, $\alpha_{2} = 1.93\pm0.28$, and the time the LC breaks between $\alpha_{1}$ and $\alpha_{2}$ being $38.0\pm7.3$~days ($\chi^2/$dof$ = 389.2/348$).  }
         \label{FigXRT}
\end{figure}

%If we force a break-time of $T_{\rm B} = 14$~days, we find $\alpha_{1} = 0.72\pm0.01$, $\alpha_{2} = 1.39\pm0.05$ ($\chi^2/dof = 394.6/349$).

\begin{figure}
   \centering % trim={<left> <lower> <right> <upper>}%, 
   \includegraphics[width=\columnwidth]{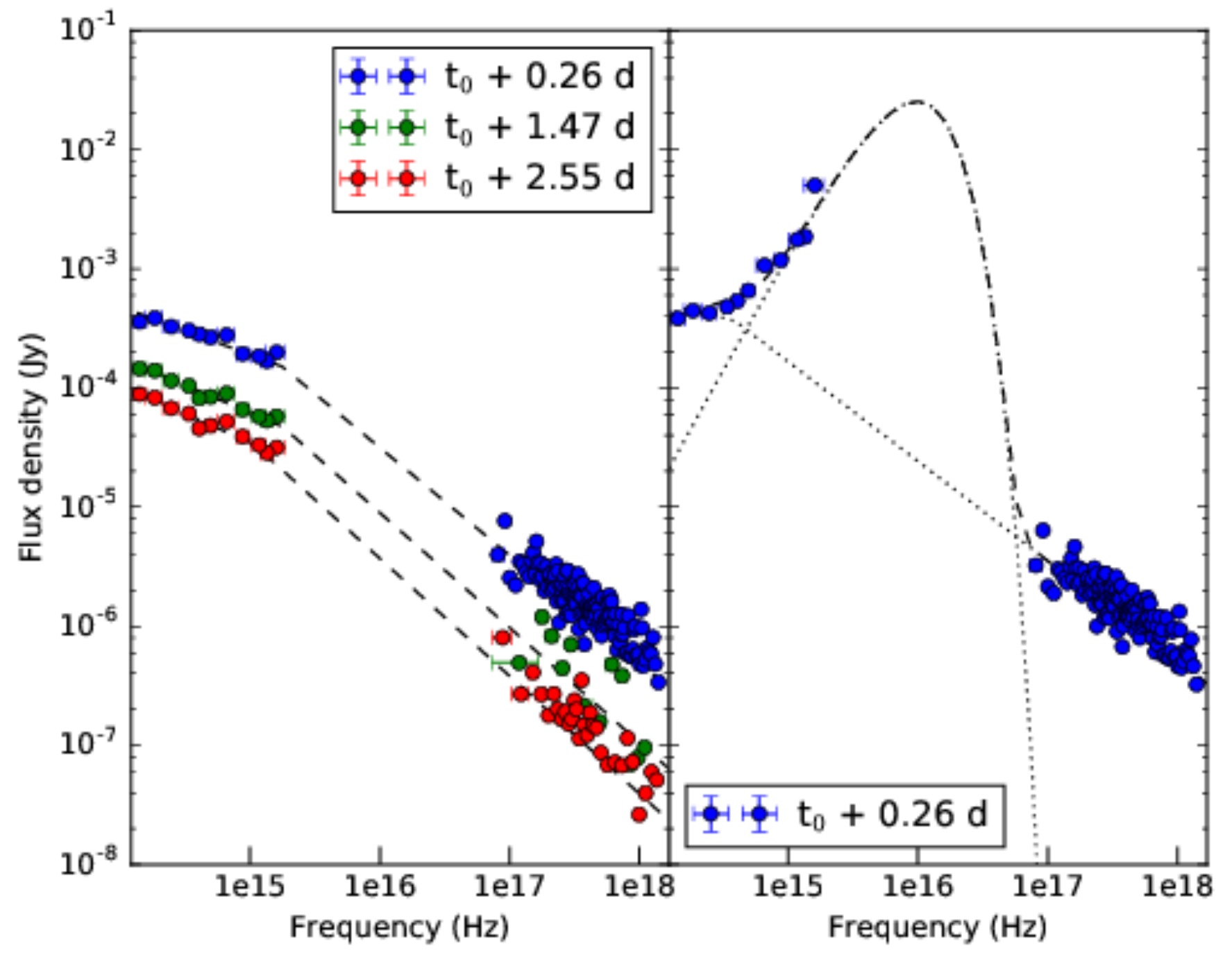}
      \caption{NIR to X-ray SED of GRB~161219B at $t-t_0 = 0.26, 1.47$ and 2.55~days.  \textit{Left}: The best-fitting model consisting of a broken power-law and MW extinction curve.  For all three epochs, we find: $\beta_{\rm opt} \approx 0.45$, $\beta_{\rm X} \approx 0.95$ and a break frequency of $\nu_{\rm B} \approx 1.7-1.8\times10^{15}$~Hz.  The local reddening is low ($E(B-V) \approx 0.02$~mag). \textit{Right}: To the SED at $+0.26$~days, we fit an extinguished broken power-law with an additional blackbody component.  At this early epoch ($\approx+5.4$~hours, rest-frame), the BB component is very hot ($T_{\rm BB} \approx 0.16\times10^{6}$~K) and it contributes $\approx 68$\% of the total recorded flux. (Note that the BB fit implies a larger local reddening than the BPL fits $-$ see the main text for further discussion).}
         \label{FigSED}
\end{figure}

We fit the \emph{Swift}-XRT X-ray LC (see Section \ref{sec:SED}) with a broken power-law (BPL; \citealt{Beuermann99}) to find the power-law decay indices $\alpha_{1}$ and $\alpha_{2}$, and the time the LC transitions ($t_{\rm B}$) between them.  Allowing all of the parameters to vary freely, our best-fitting results are: $\alpha_{1} = 0.79\pm0.02$, $\alpha_{2} = 1.93\pm0.28$, and $t_{\rm B} = 38.0\pm7.3$~days ($\chi^2/$dof$ = 389.2/348$).  Note that we excluded all data before $t-t_{0} = 0.05$~days due to the presence of an early flare, which peaked at roughly 400~s after the first detection of the GRB.  The data and best-fitting model are presented in Fig.~\ref{FigXRT}.

The rest-frame break-time measured here ($33.1\pm6.4$~days) is at a much later time than that determined by \citet{Ashall17}, who found a break-time of $\approx 12$ days ($\approx 13.8$~days observer-frame), fit over a shorter time interval (up to +30~days).  We note that if we force a break-time of $t_{\rm B} = 14$~days, we obtain decay indices of $\alpha_{1} = 0.72\pm0.01$, $\alpha_{2} = 1.39\pm0.05$ ($\chi^2/$dof$ = 394.6/349$).  %Both goodness-of-fit statistics are comparable.

\begin{figure*}
   %\centering
   \includegraphics[scale=0.417]{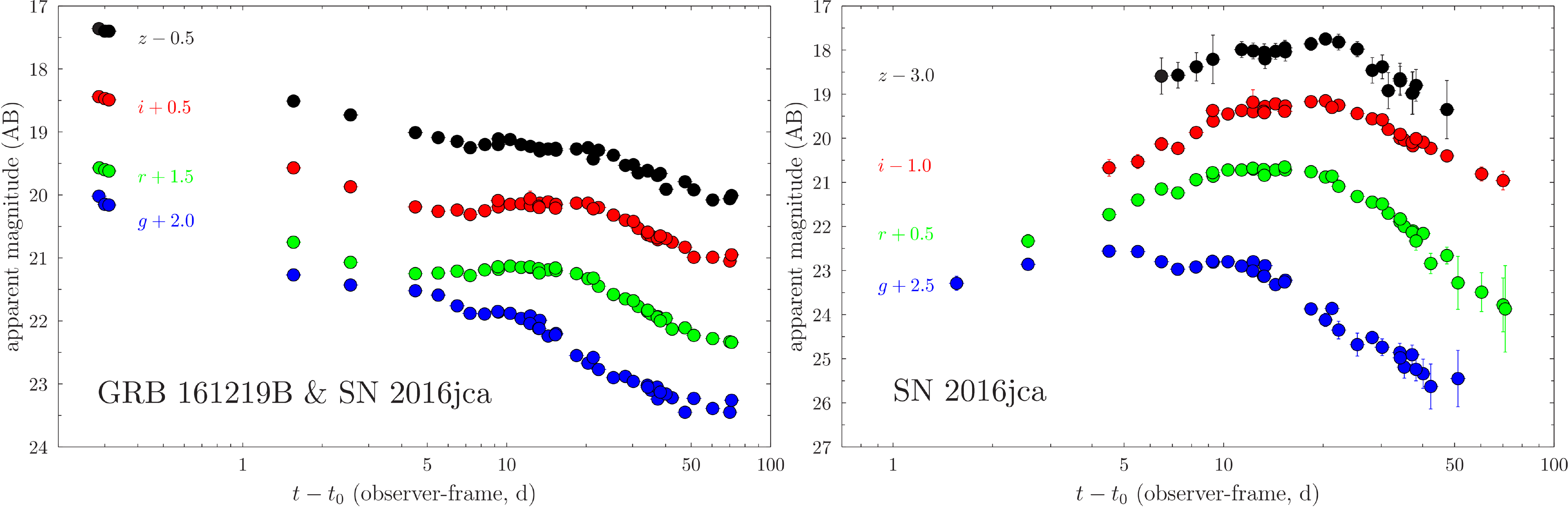}
      \caption{Observer-frame optical ($griz$) LCs of GRB~161219B / SN~2016jca. \textit{Left}: LCs of the AG, SN and underlying host, which are uncorrected for extinction.  The typical evolution from the AG-dominated to the SN-dominated phase is seen in all optical filters. \textit{Right}: Host- and AG-subtracted LCs of SN 2016jca. A pre-maximum bump is seen in the $g$-band, which peaks around $\approx 5-6$~days. This bump is conspicuously absent in the other filters, and the origin of this flux excess is discussed in Sect. \ref{sec:gbandLC}.  In contrast, the potential bump seen in the $z$-band LC around $10$~days is not real, but arises from instrumental defects.}
    \label{FigLC_optical}
   \end{figure*}

\begin{table*}
%\scriptsize
\centering
\setlength{\tabcolsep}{10.0pt}
\setlength{\extrarowheight}{3pt}
\caption{AG \& SN phenomenological properties}
\label{tableGRB_SN_obs_props}
\begin{tabular}{cccccccc}
\hline																					
Band	&	$\alpha_{1}$ & $\alpha_{2}$ & $t_{\rm B}$ (days) 	&	$k$	&	$s$	&	$m_{\rm p}$ (mag)	&	$t_{\rm p}$ (days)	\\
\hline
X-ray	&	$0.79\pm0.02$	&	$1.93\pm0.28$	&	$38.0\pm7.30$	&	-	&	-	&	-	&	-	\\
X-ray	&	$0.72\pm0.01$	&	$1.39\pm0.05$	&	$14.0^{\dagger}$	&	-	&	-	&	-	&	-	\\
$g$	&	$0.86\pm0.06$	&	-	&	-	&	$0.77\pm0.08$	&	$0.58\pm0.05$	&	$20.31\pm0.05$	&	$9.5\pm1.0$	\\
$r$	&	$0.80\pm0.05$	&	-	&	-	&	$0.78\pm0.09$	&	$0.78\pm0.03$	&	$20.18\pm0.05$	&	$14.1\pm1.0$	\\
$i$	&	$0.76\pm0.07$	&	-	&	-	&	$0.81\pm0.06$	&	$0.89\pm0.03$	&	$20.20\pm0.04$	&	$17.4\pm1.1$	\\
$z$	&	$0.75\pm0.04$	&	-	&	-	&	$0.51\pm0.06$	&	$0.64\pm0.03$	&	$20.90\pm0.05$	&	$17.8\pm0.9$	\\
$J$	&	$0.61\pm0.05$	&	-	&	-	&	$0.83\pm0.12$	&	$0.87\pm0.12$	&	$19.54\pm0.10$	&	$25.8\pm2.5$	\\
$H$	&	$0.59\pm0.05$	&	-	&	-	&	-	&	-	&	-	&	-	\\
$K$	&	$0.63\pm0.26$	&	-	&	-	&	-	&	-	&	-	&	-	\\
\hline															
$B$	&	$0.76\pm0.04$	&	-	&	-	&	$0.65\pm0.09$	&	$0.57\pm0.12$	&	$20.52\pm0.05$	&	$7.9\pm1.3$	\\
$V$	&	$0.79\pm0.06$	&	-	&	-	&	$0.79\pm0.09$	&	$0.79\pm0.11$	&	$20.18\pm0.05$	&	$12.3\pm0.7$	\\
$R$	&	$0.76\pm0.04$	&	-	&	-	&	$0.80\pm0.12$	&	$0.89\pm0.09$	&	$20.02\pm0.06$	&	$15.8\pm1.7$	\\
\hline																					
\end{tabular}
%\medskip
\begin{flushleft}
$\dagger$ Break-time fixed during fit.\\
NB: X-ray data and filters $grizJHK$ are for observer-frame filters and times. Rest-frame properties are given for $BVR$.\\
NB: Filters $griz$ are in the AB system, while $BVR$ and $JHK$ are in the Vega system.\\
NB: Properties in filters $griz$ and $BVR$ have been host-subtracted, whereas those in $JHK$ are not. \\
\end{flushleft}
\end{table*}

\begin{figure*}
   \centering
   \includegraphics[scale=0.93]{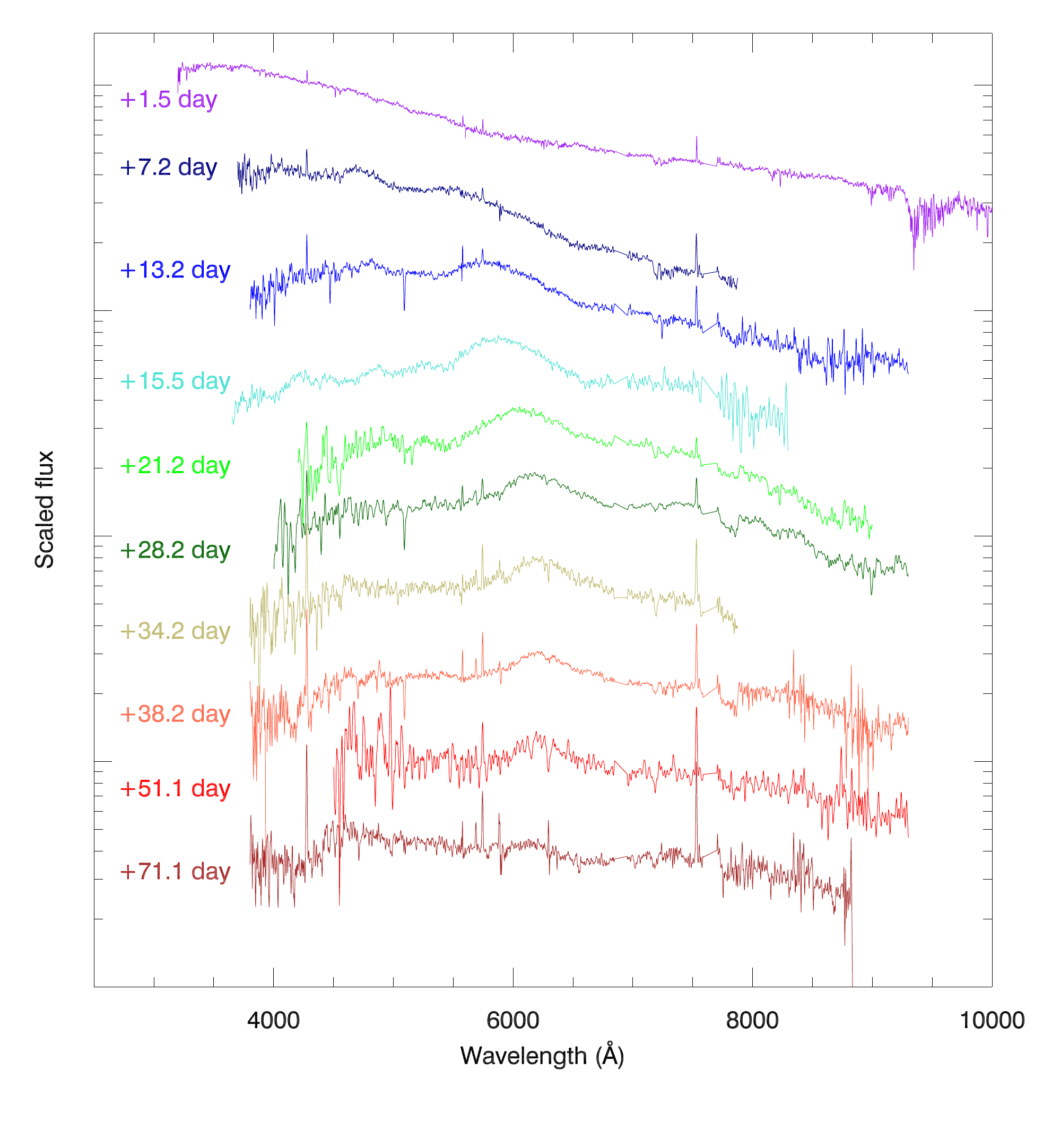}
      \caption{Spectroscopic time-series of GRB 161219B \& SN 2016jca.  Wavelengths and times are presented in the observer-frame ($z=0.1475$), and are not host-subtracted.  Narrow emission lines are seen, which arise from star-forming regions within the host galaxy.}
   \label{FigSpectra}
\end{figure*}

%Interestingly, an excess of flux is seen in the early $g$-band LC during $\approx 3-6$~days.

\section{The Spectral Energy Distribution}
\label{sec:SED}

We modelled the spectral energy distribution (SED) from NIR to X-ray frequencies in order to determine any dust extinction in the GRB's local environment. We calculated flux densities from the GROND (AB) magnitudes fluxes complemented by flux densities calculated from the \emph{Swift}-UVOT AB magnitudes from optical to near-UV wavelengths, and \emph{Swift}-XRT data. We computed SEDs at three different epochs, whose mean times were chosen with respect to \emph{Swift}-XRT data: 1) $t-t_0 = 0.26$~days; 2) $t -t_0 = 1.47$~days; and 3) $t -t_0 = 2.55$~days. The XRT data were reduced with the standard \texttt{XRTPIPELINE} tool in \texttt{caldb} (version 20160609), after which we extracted time-resolved spectra that corresponded to the three different SED epochs using \texttt{XSELECT}. For the first SED, data suffered from pile-up and, therefore, we used an annular region with an inner radius of 5~pixels and an outer one of 30~pixels, while for the remaining two SEDs we used a circular region with a radius of 20~pixels.

UVOT flux densities were computed using standard \emph{Swift}-UVOT data-reduction procedures (\texttt{caldb} version 20170130). First, we determined the regions for the source and background from the summed UVOT-$v$-band image, using circular apertures with a radius of 6$^{\prime\prime}$ for both regions. Then we used the \texttt{UVOTMAGHIST} tool on the level~2 fits images to determine AB flux densities in each filter.  Next, using the foreground-corrected flux densities of the AG+SN+host determined from the GROND observations, we fit a SPL to the LCs and interpolated the flux densities to the time of the second and third SEDs.  The flux densities for the first epoch, which occurred before the first GROND epoch ($+0.286$~days), were determined by extrapolating the SPL to the time of the first SED.  All \emph{Swift}-UVOT flux densities were determined via LC interpolation as the UVOT magnitudes were obtain before and after the times of the chosen SED epochs.  Note that the UVOT magnitudes/fluxes are contaminated with host flux: as no pre-explosion UVOT templates of the GRB exist, we were unable to remove this component from the analysis.  As such, in order to provide a consistent analysis, all fluxes modelled in this section have not had the host flux removed.

The final NIR-to-Xray SEDs were fitted using \texttt{XSPEC} \citep{Arnaud96}.  During the fit (following the general procedure described in \citealt{Schady10}), we included two dust and gas components corresponding to the Galactic (using the UV/optical/NIR extinction law fom \citealt{Cardelli89}) and host galaxy photoelectric absorption \citep{Wilms2000} and dust extinction (using the MW/SMC/LMC templates from \citealt{Pei92}), where we fixed the Galactic values to $E(B-V) = 0.028$ mag and $N(H\textsc{i}) = 3.06 \times 10^{20}$ cm$^{-2}$ \citep{Willingale13}. For the Galactic and host X-ray absorption component, we used the Tuebingen-Boulder ISM absorption model.  For the fit, and in addition to the three dust extinction templates, we tried several different scenarios: (1) a SPL with no break between the X-ray and UV/optical/NIR regimes (e.g. \citealt{Zafar11}); (2) a BPL, where the cooling break occurs between these regimes (i.e. $\beta_{\rm X} = \beta_{\rm opt} + 0.5$); (3) a BPL with the break frequency between the optical and X-ray, but we allowed the spectral indices to vary freely with no constraints.

We used the first epoch at $t-t_0 = 0.26$~days to determine which of these scenarios provided the best fit.  First, the simple absorbed SPL function resulted in a goodness-of-fit of $\chi^2/$dof$ = 662.6/573$, with a spectral index of $\beta = 0.77\pm0.02$. Next, for scenario (2), we found spectral indices of $\beta_{\rm opt} = 0.40 \pm 0.07$ and $\beta_{\rm X} = 0.90 \pm 0.07$, with a break frequency of $\nu_{\rm B} = (1.76\pm1.91)\times10^{15}$~Hz ($\chi^2/$dof$ = 433.2/573$), which is just outside of the UVOT frequency range (i.e. in the UV).  The reddening was found to be $E(B-V)=0.026\pm0.020$~mag, with identical values found (within their respective errorbars) for all three dust extinction templates.  For scenario (3), where the spectral indices were allowed to vary, we found similar values for the free parameters: $\beta_{\rm opt} = 0.44 \pm 0.08$ and $\beta_{\rm X} = 0.89 \pm 0.09$, with a break frequency of $\nu_{\rm B} = (1.71\pm1.92)\times10^{15}$~Hz, and an extinction of $E(B-V)=0.016\pm0.023$~mag ($\chi^2/$dof$ = 433.3/572$).  These results suggest that there is very little extinction local to the GRB.  

Although the $\chi^2$ statistic is more favourable for scenario (1), we ruled against this model for two reasons: first, when fitting the optical and X-ray regimes separately, we found very different spectral indices of $\beta_{\rm opt} = 1.10\pm0.07$ ($\chi^2/$dof$ = 8.1/7$) and $\beta_{\rm X} = 1.98\pm0.10$ ($\chi^2/$dof$ = 356.0/566$).  Secondly, an $F$-test between scenarios (1) and (3) show that the latter scenario (i.e. the BPL) is favoured: the $F$-value is $F=302.7$, with a probability of $9.9\times10^{-55}$.  We therefore conclude that the NIR to X-ray SED at $t-t_0 = 0.26$~days is best-fit with an extinguished BPL, which is similar to the SED modelling results of \citet{Ashall17}.

For the other two epochs at $t -t_0 = 1.47$~days and $2.55$~days we found similar results for all free parameters.  The optical and X-ray indices do not vary much between $0.26-2.55$~days, which have an average value of $\beta_{\rm opt} \approx 0.45$ and $\beta_{\rm X} \approx 0.95$, while the break frequency is approximately $\nu_{\rm B} \approx 1.7-1.8\times10^{15}$~Hz.  The error-bars on the break frequency are too large to determine if it increases or decreases with time.  Finally, the weighted average of the line-of-sight host extinction is $E(B-V)_{\rm host, weighted} = 0.017\pm0.012$~mag, which is the value used throughout this paper.

\begin{figure}
   %\centering
   \includegraphics[width=\columnwidth]{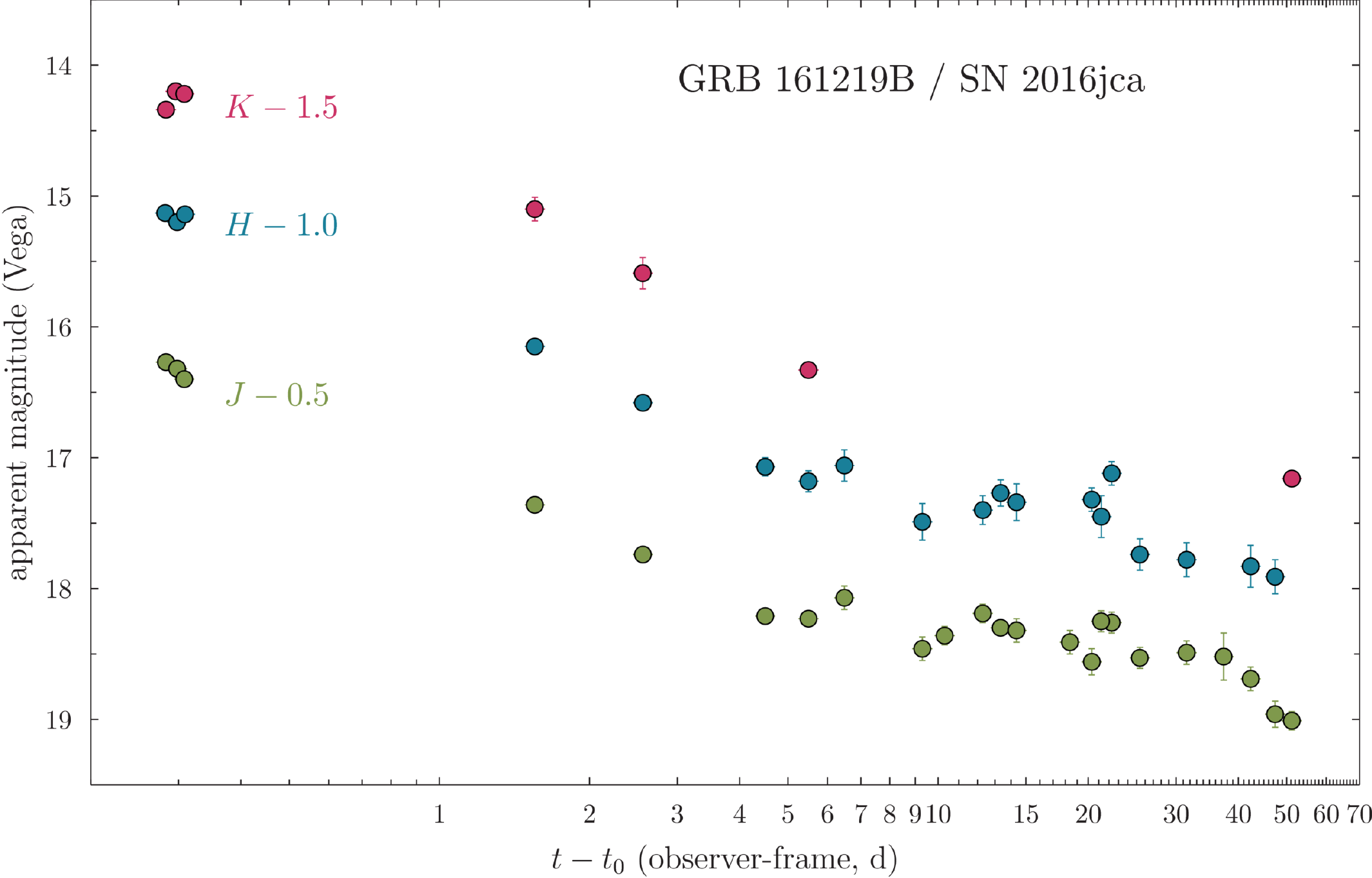}
      \caption{Observer-frame NIR ($JHK$) LCs of GRB~161219B / SN~2016jca.  The magnitudes are not corrected for extinction, and they have an unknown contribution from the underlying host. }
         \label{FigLC_IR}
\end{figure}

\subsection{An Extra Blackbody Component?}

Motivated by previous studies that found an extra thermal component in the early X-ray spectra of many GRBs (e.g. \citealt{Campana06,Starling11,Page11,Thoene11,SparStar12,FriisWatson13,Schulze14}, we also performed a fourth fit where we included an extra blackbody (BB) component.  For the first epoch, we found a BB component of temperature $T_{\rm BB} \approx 0.16\times10^6$~K and radius $R_{\rm BB} \approx 6\times10^{14}$~cm ($\chi^2/$dof$ = 404.3/578$).  An $F$-test between this model and scenario 3 gives an $F$-value of 20.44 and a probability of $2.7\times10^{-9}$, indicating it may provide a better fit to the data despite the increase in free parameters.  In relative terms, the fit suggests that the BB component contributes 68\% of the total flux at this epoch.  We also find that the fit gives a larger rest-frame reddening of $E(B-V)=0.16\pm0.13$~mag.

We fit the latter two epochs with the same fit.  For the second epoch, we found that the fractional contribution of the BB component decreased to $\approx10$\%.  The BB fit to the third epoch was entirely unconstrained.  For the second epoch, we found the temperature decreased to $T_{\rm BB} \approx 32,000$~K, while the BB radius was roughly the same as the first epoch.  We note that the error-bars are too large to infer any changes/evolution.  Again, the fit suggests larger host-extinction of $E(B-V)\approx 0.15$~mag, but it is very poorly constrained.  %We also find  %fdgsxfdggfdgdfgffdgfdgfdgfdgsgfgfdgfdb fdghfghgfh fghfghgfh gfhgfhghf gfhgf hgfhgf

If we take these results at face-value, the cooling thermal component could imply the presence of a thermal cocoon surrounding the jet, which is very hot early on, but fades rapidly, and does not contribute any appreciable flux after a couple days (rest-frame). However, the cocoon radius estimated for GRB~130925A by \citet{Piro14} is of order $0.4-1.4\times10^{11}$~cm, which is more than three orders of magnitude smaller than the radius found here. Alternatively, the thermal component could arise from a scenario similar to that suggested by \citet{Campana06} for GRB~060218, where the shock breakout was trapped in an optically-thick stellar wind, breaking out only after the wind became optically thin.  A consequence of this model are the pre-maximum peaks seen in the optical and UV LCs of GRB~060218, which are not seen here. We note that the alternative scenario presented by \citet{Margutti15} and \citet{Nakar15}, where the breakout occurs from a low-mass, extended envelope surrounding the progenitor, also predicts the achromatic pre-maximum bumps.  Regardless, \cite{Campana06} found that the BB radius of the thermal component evolved from $\approx5\times10^{11}$~cm at +300~s to $\approx3\times10^{14}$~cm at 0.9~days.  At +0.3~days it had a radius of $\approx10^{13}$~cm, almost two orders of magnitude smaller than that found here.  Next, \citet{Starling11} found a radius of $\approx 8\times10^{11}$~cm ($t-t_0<800$~s) for GRB~100316D.  For the sample of LGRBs presented in \citet{Starling12}, the BB radii determined from early-time X-ray spectra ($t-t_0=80-800$~s), range from $0.03-9\times10^{12}$~cm.  A similar range of radii was determined by \citet{Page11} for GRB~090618 for early-time X-ray spectra.

Interestingly, a BB component was found in a fit of the X-ray to NIR SEDs of GRB~120422A by \citet{Schulze14}, where at $+0.267$~days, they find a BB temperature of $T_{\rm BB} \approx 0.19\times10^{6}$~K, and a radius of $R_{\rm BB} \approx 7\times10^{13}$~cm, which is about one order of magnitude smaller than that found here.  This BB component was interpreted as thermal emission arising from the cooling stellar envelope following shock breakout, and a similar interpretation of the thermal component for GRB~161219B is appealing.  Note that pre-maximum bumps were not observed for GRB~120422A, similar to GRB~161219B.

%As additional free parameters can inevitably lead to better fits of model to data, we checked if the BPL+BB model could be considered statistically significant with respect to the freely varying BPL model by performing an F-test within \texttt{XSPEC}. We found an F-value of 13.6 ($p < 0.0005$), which for the number of degrees-of-freedom in each model (BPL: 573; BPL+BB: 570), in well in excess the critical value, confirms that there was a significant effect in using the BPL+BB model.

\begin{figure}
   %\centering
   \includegraphics[width=\columnwidth]{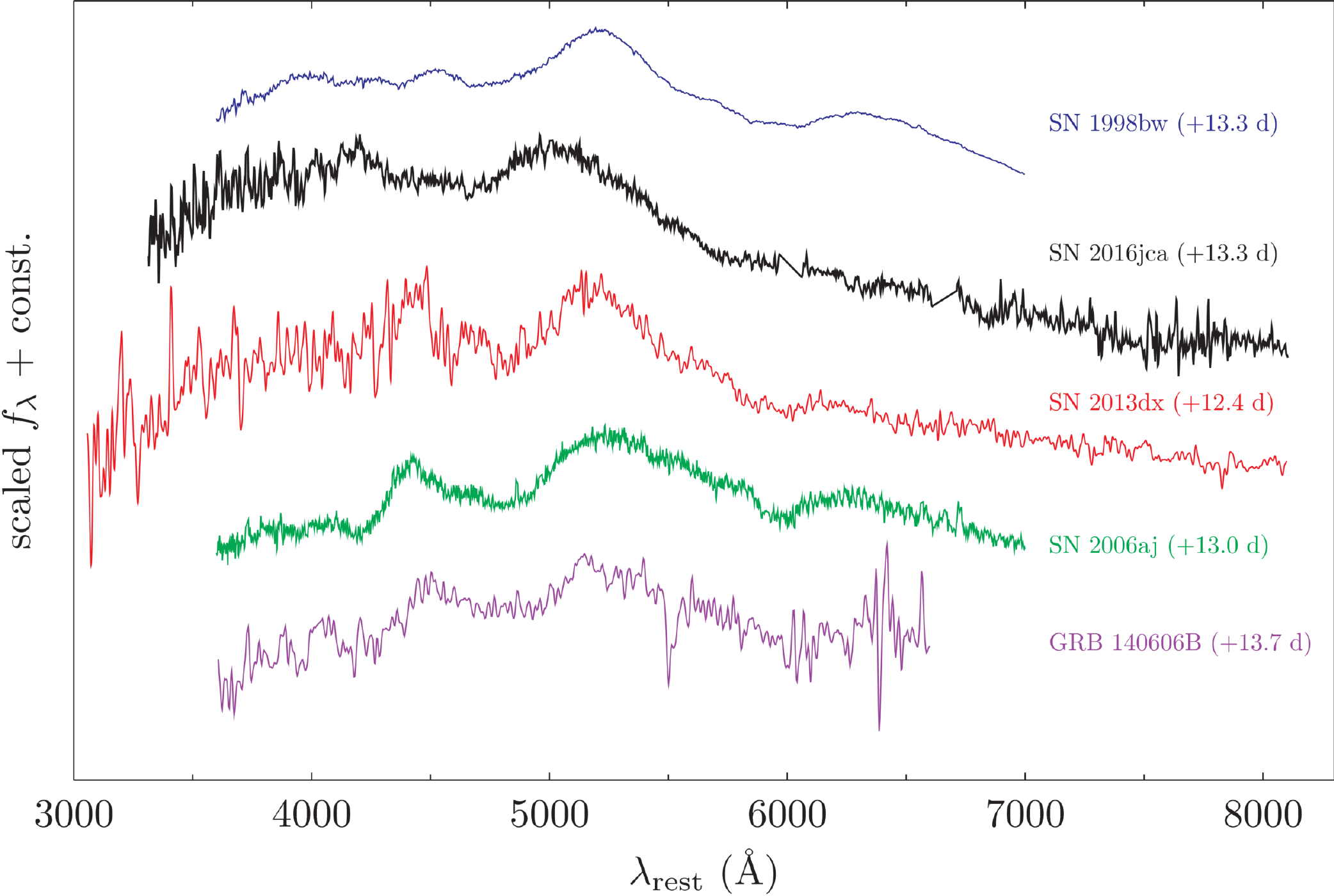}
      \caption{Comparison of the spectrum of SN 2016jca obtained with the GTC on 01-January-2017 ($t-t_0 = 13.3$~days; shown in black), which was obtained near peak $V$-band light ($+12.5$~d, rest-frame).  Plotted for comparison is a sample of GRB-SNe at similar post-explosion times: SN~1998bw ($+13.3$~days; blue), SN~2006aj ($+13.0$~days; green), SN~2013dx ($+12.4$~days; red) and the SN associated with GRB~140606B ($+13.7$~days; purple).  All times and wavelengths are shown in the rest-frame.  Visual inspection of the spectra reveals that the trough bluewards of the peak around $5000\sim5200$~\AA, which we attribute to blueshifted Fe~\textsc{ii} $\lambda5169$, occurs at bluer wavelengths for SN~2016jca than all the comparison GRB-SNe, thus highlighting its high-velocity nature.}
   \label{Figspeccompare}
\end{figure}

\section{The Afterglow}
\label{sec:AG}

The observer-frame optical ($griz$) and NIR ($JHK$) LCs of GRB~161219B/SN~2016jca are shown in Fig. \ref{FigLC_optical} and Fig. \ref{FigLC_IR}.  

To quantify the three sources of flux portrayed in the LCs (the AG, the SN and the host galaxy, e.g. \citealt{Zeh04}), we first de-reddened the observations of the OT for foreground extinction (see Section \ref{sec:SED} and Table \ref{table:GRB_vitals}).  We then converted the magnitudes into monochromatic flux densities using the AB zeropoint flux density (for $griz$) and the flux density zeropoints for $JHK$ from \cite{Greiner08}.  The foreground de-reddened host flux densities in each optical filter were subtracted, resulting in LCs of just the optical transient (Fig. \ref{FigLC_optical}), which were then corrected for host/local extinction.  Unfortunately, pre-explosion images of the host in the NIR filters $JHK$ do no exist (they are too shallow in the 2MASS survey), and hence the NIR data in Fig. \ref{FigLC_IR} are de-reddened but not host subtracted.

In order to quantify the AG component we fit both a SPL and a BPL to the optical/NIR LCs.  In this study, we did not assume that the AG behaves achromatically (e.g. \citealt{Klose04,Kann16}).  In filters $griz$ we also included a template SN in the fit (SN~1998bw), and simultaneously determined how much SN~2016jca was brighter/fainter ($k$) and wider/slimmer ($s$) than the template (see Section \ref{sec:SN_optical}).  The fitting was done using scripts written in \texttt{pyxplot}, as employed in previous works \citep{Cano2011a,Cano2011b,Cano+14,Cano15}, which use linear-least squares fitting to find the best-fitting values of each one of these free parameters. Our results are presented in Table \ref{tableGRB_SN_obs_props}.

It can be seen that the assumption of achromatic AG behaviour in other studies would also be justified in this case.  For all filters, a SPL provided the best fit to the optical observations, where in all filters the decay index was $\alpha \approx 0.8$ in $griz$ and in the NIR it was $\alpha\approx 0.6$.  In comparison, the decay index at times $<40$~days in the X-ray regime was found to be $\alpha_{\rm X}=0.79\pm0.02$, in excellent agreement with that found in the optical filters.  One caveat to the fitting is that the host-contribution was only removed from the optical observations and not the NIR, where pre-explosion imaging in $JHK$ is not available.  The effect of having the host flux in the NIR LCs is that the decay rate will be slower than in reality: as the AG fades, the host contributes an increasing portion of flux to the LC.  Thus the AG will (incorrectly) appear to fade at a slower rate, and hence have a smaller value of $\alpha$.

\begin{figure}
   %\centering
   \includegraphics[width=\columnwidth]{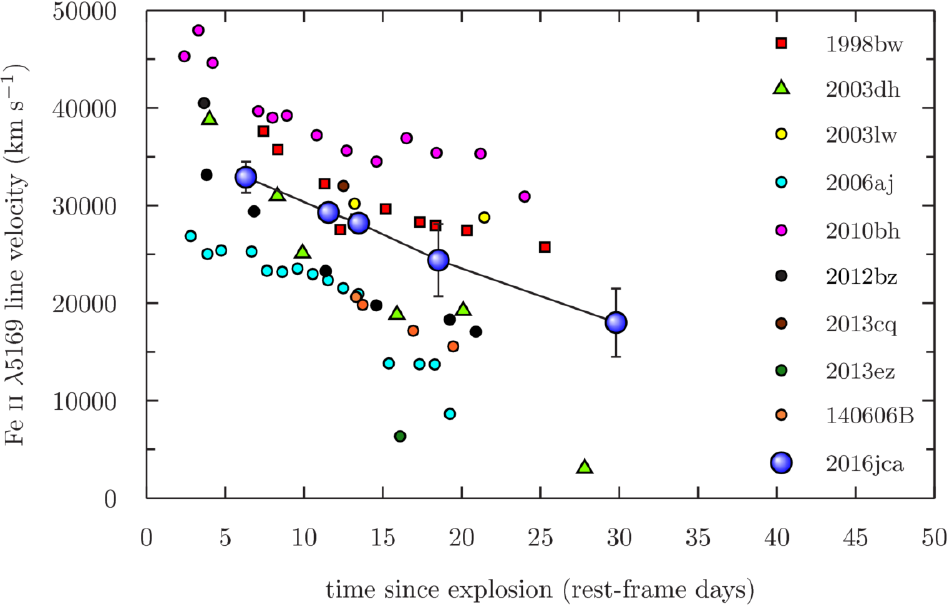}
      \caption{Blue-shifted velocities of the Fe \textsc{ii} $\lambda$5169 transition.}
   \label{FigLineVel}
\end{figure}

\section{SN 2016jca - Observational Properties}
\label{sec:SN_optical}

\subsection{Photometric Properties}
\label{sec:SN_optical_photo}

As well as determining the AG behaviour in Section \ref{sec:AG}, we simultaneously fit the AG and SN to determine the luminosity, $k$, and stretch, $s$, factors of SN~2016jca relative to SN~1998bw.  The luminosity factor is similar in optical filters $gri$, with a value of $k \approx 0.8$, but it is fainter in the $z$-band ($k=0.5$).  The stretch factors in filters $g$ and $z$ are $s\approx0.6$, while in $riJ$ they are $s=0.8-0.9$.  Collectively, these results show that in all filters $gri$, the SN is fainter and evolves more quickly than SN~1998bw.

We also fit the AG- and host-subtracted SN LCs with two models: (1) a model based on the equations in \citet{Bazin11}, and (2) high-order polynomials, in order to determine the time and magnitude of maximum light in each filter.  These are also given in Table \ref{tableGRB_SN_obs_props}.  It is seen that the SN peaks later in redder filters, as expected.

Finally, we fit the rest-frame SN LCs (i.e. K-corrected; see Sect. \ref{secksrelation}) in $BVR$ to determine their peak magnitudes and time of peak light.  As with the observer-frame filters, the SN peaks at later times in redder filters.  Relative to SN~1998bw, SN~2016jca was fainter ($k=0.65-0.80$), and reached peak light before SN~1998bw ($s=0.57-0.89$) in all filters.

Next, and using a distance modulus of $\mu = 39.22$~mag, we find rest-frame, peak absolute magnitudes of $M_{B} = -18.70\pm0.05$, $M_{V} = -19.04\pm0.05$ and $M_{R} = -19.20\pm0.06$.  For comparison, \citet{LiHjorth2014} found for SN~1998bw a peak $V$-band absolute magnitude of $M_{V}=-19.3$.  Thus SN~2016jca is roughly 0.25~mag fainter than the archetype GRB-SN~1998bw, which agrees with the fact that the luminosity factor in this filter ($k_{V} = 0.79$) is less than one.  Relative to the rest-frame magnitudes of the GRB-SN sample in \citet{LiHjorth2014}, SN~2016jca is quite faint, and is only brighter than SN~2006aj ($M_{V} = -18.85$) and SN~2010bh ($M_{V} = -18.89$).

\begin{figure}
   %\centering
   \includegraphics[width=\columnwidth]{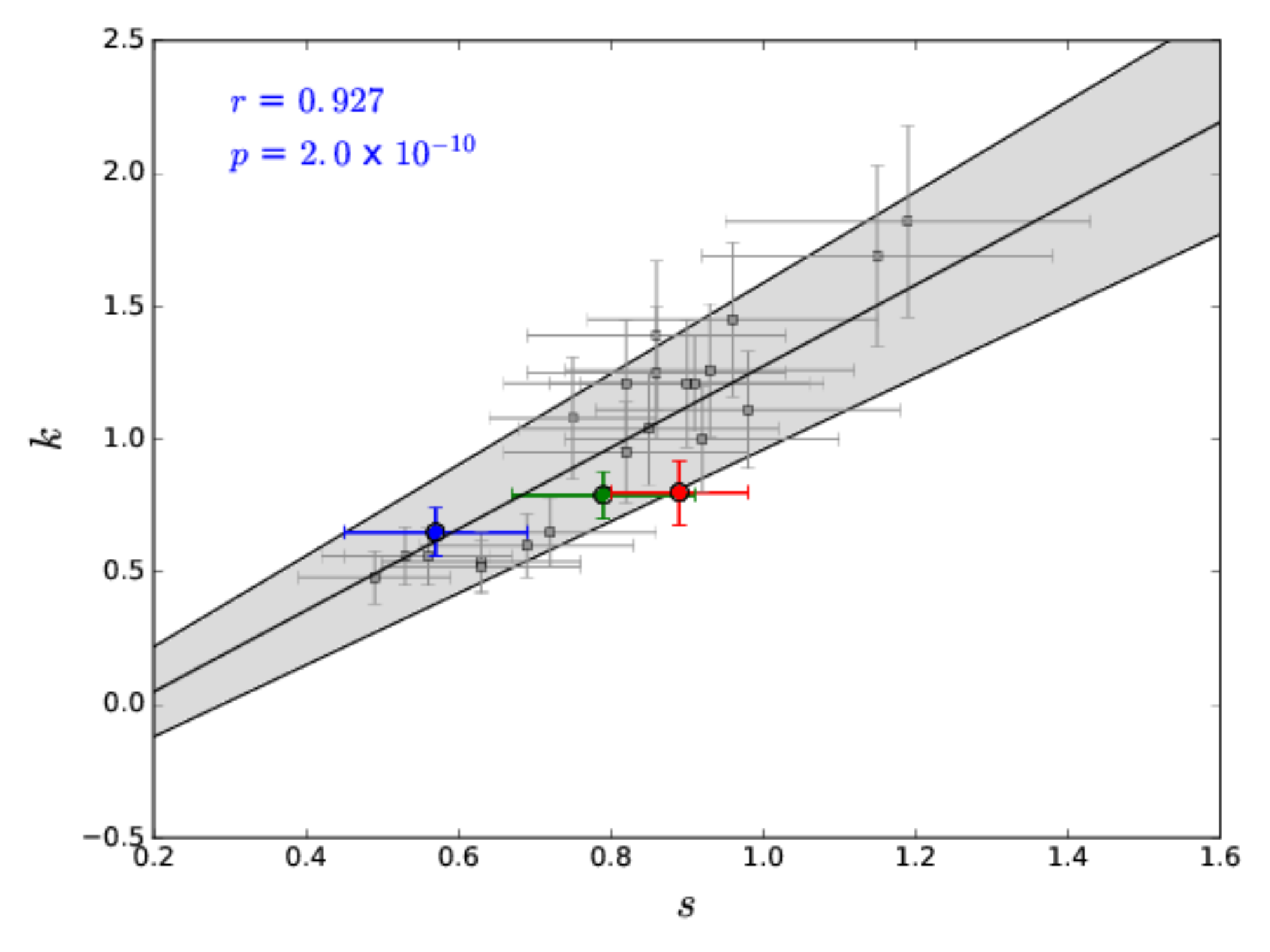}
      \caption{Luminosity ($k$)-stretch ($s$) relationship from \citet{Cano14}.  Plotted also are the rest-frame $s$ and $k$ values of SN 2016jca in $B$ (blue), $V$ (green) and $R$ (red).  The best-fitting is shown in black, and the $1\sigma$ error region is shown in shaded grey.  The best-fitting values for the fitted line are found in the main text. The $k$ and $s$ values in $R$ are marginally consistent within $1\sigma$, while those in $B$ and $V$ are entirely consistent with the relationship. }
   \label{Figks}
\end{figure}

\subsection{Spectroscopic Properties}

Our spectroscopic time-series of GRB~161219B / SN~2016jca is shown in Fig. \ref{FigSpectra}.  The transition from the AG-dominated to the SN-dominated phase is clearly portrayed in the shape and evolution of the optical spectra.  The XS spectrum taken at $t-t_0=1.3$~days (rest-frame) is flat and featureless, typical of GRB afterglow spectra (e.g. \citealt{Fynbo09}).  Host emission lines are superimposed upon the AG spectrum.  The next spectrum, obtained at $t-t_0=6.3$~days (rest-frame) shows unambiguous spectral features of a broad-lined SN, while the absence of both hydrogen and helium indicates a spectral class of type Ic. The broad absorption feature seen at observer-frame 5200~\AA~is attributed to blue-shifted Fe \textsc{ii} $\lambda$5169, though other transitions may also be blended in.  Near and after peak light, an additional absorption feature is seen near observer-frame 6500~\AA, which may be blueshifted Si \textsc{ii} $\lambda$6355.  A hint of blueshifted O \textsc{i} and/or Ca \textsc{ii} near observer-frame 9000~\AA~is seen in the GTC spectrum taken on 16-January-2017 ($t-t_0=28.2$~days, observer-frame).  Sky lines in the 22-January-2017 spectrum ($t-t_0=34.2$~days, observer-frame) inhibit our ability to detect the same feature.

A comparison of our GTC spectrum taken near peak $V$-band light  (01 January, 2017) with other GRB-SNe near peak $V$-band light is shown in Fig. \ref{Figspeccompare}.  The broad spectral features seen for SN~2016jca are quite typical of other GRB-SNe.   In Fig. \ref{FigLineVel} we plotted the blueshifted velocities of Fe~\textsc{ii} $\lambda$5169, under the assumption that the absorption feature at observer-frame 5200~\AA~is unblended with other transitions.  It is seen that the magnitude and evolution of the Fe~\textsc{ii} $\lambda$5169 is quite typical of other GRB-SNe.  At peak bolometric light ($t-t_0 = 10.7$~days, rest-frame; see Sect. \ref{sec:bolo}), the line velocity is $v_{\rm Fe} = 29\,700 \pm 1500$~km~s$^{-1}$.  As there were no data at the precise time of peak bolometric light, we determined the peak velocity by fitting a log-linear spline to the line velocity data, and extracted the velocity at the time of peak light. In comparison, \citet{Ashall17} found a peak photospheric velocity of $\approx 26\,000$~km~s$^{-1}$ from their spectral modelling.  While good agreement is seen between the two analyses, we must consider the limitations of using a single transition as a proxy for the photospheric velocity (e.g. \citealt{Modjaz16}).

\begin{figure}
   %\centering
   \includegraphics[width=\columnwidth]{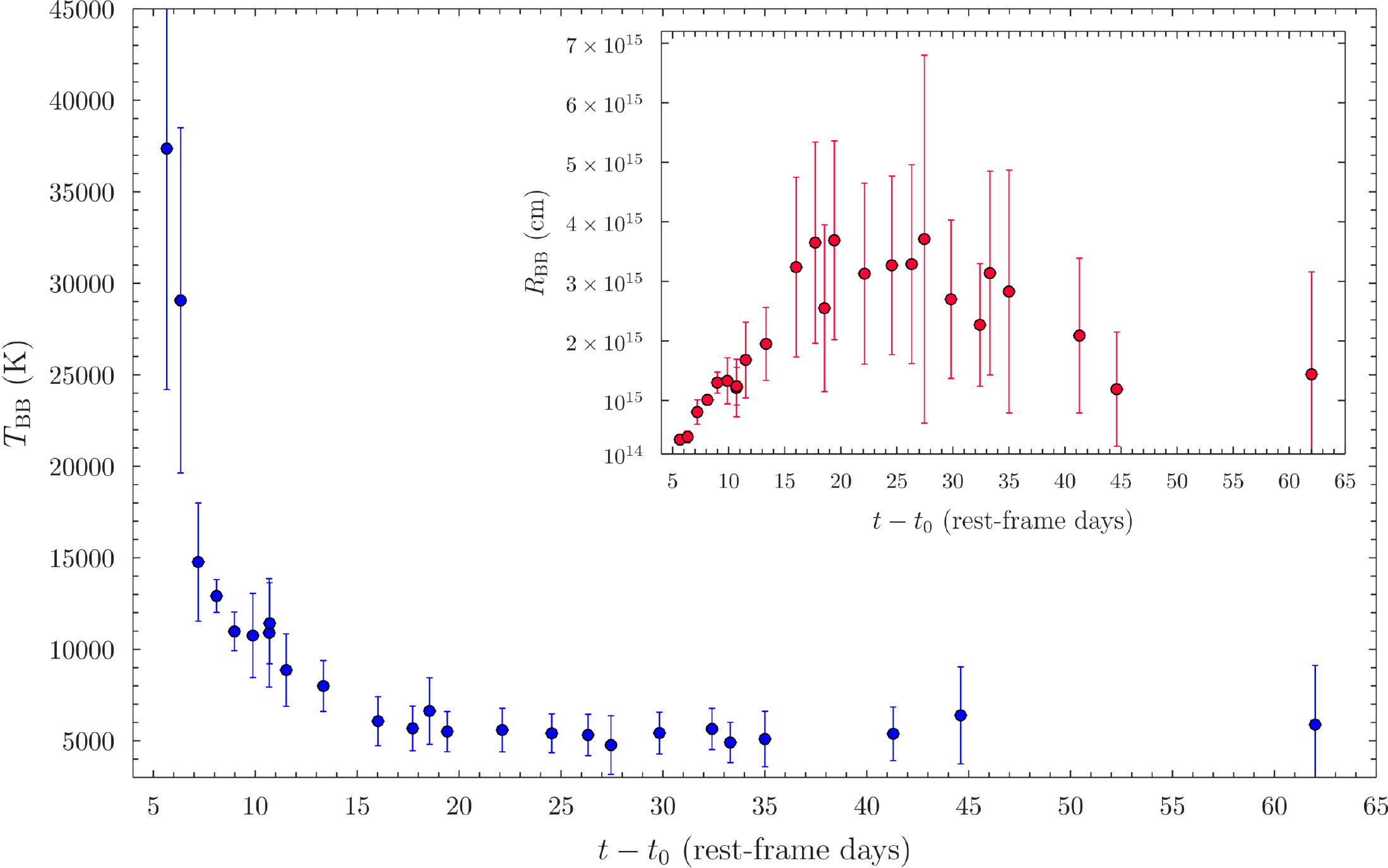}
      \caption{Evolution of the BB colour temperature ($T_{\rm BB}$, blue) and radius (inset, red, $R_{\rm BB}$).  The colour temperature corresponds to filters $griz$.}
   \label{Figtemp}
\end{figure}

\subsection{Luminosity-Stretch Relationship}
\label{secksrelation}

In an identical analysis as presented in \citet{Cano14}, we investigated the \emph{rest-frame} stretch ($s$) and luminosity ($k$) factors of SN~2016jca.  The detailed procedure is described in \citet{Cano14}, but briefly the main steps include: (1) create host-subtracted and de-reddened observer-frame SEDs in filters $griz$ for each contemporaneous epoch.  Then, using a redshift of $z=0.1475$, we interpolated to the precise red-shifted rest-frame wavelength in filters $BVR$, using the effective wavelengths from \citet{Fukugita95}.  As such, our K-correction is obtained via SED interpolation.  We then extracted the SED-interpolated flux at the desired red-shifted wavelength, recreating a LC similar to that shown in Fig. \ref{FigLC_optical}.  Next, a SPL and template SN LC were simultaneously fit to the observations to obtain the decay index $\alpha$, as well as $s$ and $k$.  The rest-frame properties are presented in Table \ref{tableGRB_SN_obs_props}.  It is seen that the decay index matches very well with those obtained from modelling the observer-frame filters.

\begin{figure*}
   %\centering% trim={<left> <lower> <right> <upper>}%, 
   \includegraphics[width=\hsize, trim={0 200pt 0 0}]{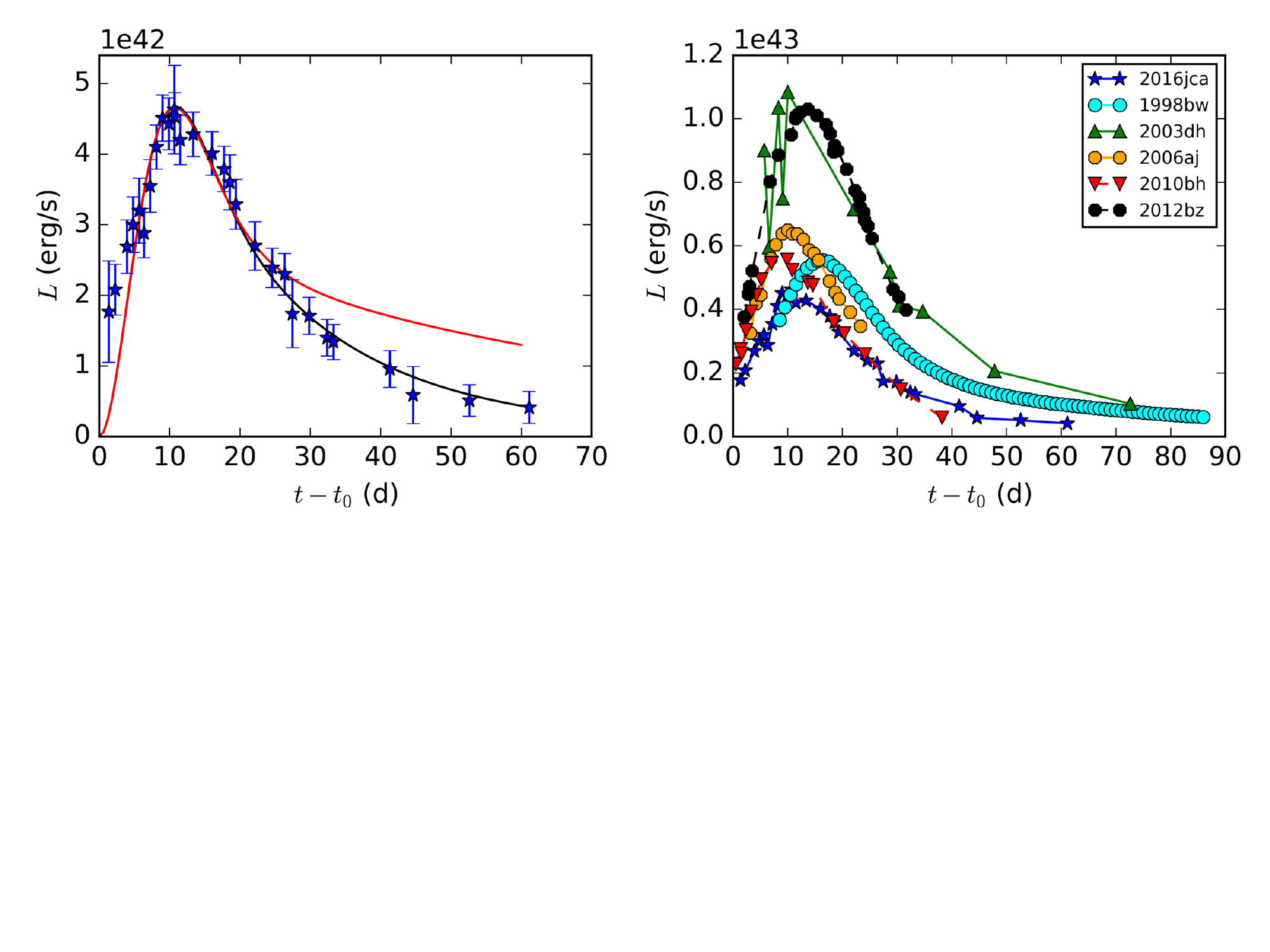}
      \caption{The radioactive heating model \citep{Arnett1982} fit to the bolometric ($griz$) LC of SN~2016jca.  \textit{Left}: Two versions of the model have been fit to the data, one that assumes full trapping of $\gamma$-rays (red), and one which allows $\gamma$-ray leakage (black).  It is seen that the latter model better fits the data.  Both models require a nickel mass of $M_{\rm Ni} = 0.22$~M$_{\odot}$, an ejecta mass of $M_{\rm ej} = 5.8$~M$_{\odot}$, and for a peak photospheric velocity of $v_{\rm ph} = 29\,700\pm1500$~km~s$^{-1}$, a kinetic energy of $E_{\rm K} = 5.1\pm0.8\times10^{52}$~erg.  Both models assume a grey optical opacity of $\kappa = 0.07$~cm$^2$~g$^{-1}$.  In the latter model, we find a $\gamma$-ray opacity of $\kappa_{\gamma} = 0.034$~cm$^2$~g$^{-1}$. \textit{Right}: Comparison of the optical bolometric LCs of a small sample of GRB-SNe: SN~1998bw ($BVRI$), SN~2003dh ($UBVR$), SN~2006aj ($BVRI$), SN~2010bh ($griz$) and SN~2012bz ($griz$).}
   \label{FigArnett}
\end{figure*}

We fit a straight line to the $k-s$ values ($N=24$, and dof$ = 24-2 = 22$), finding a slope of $m=1.53\pm0.18$ and $y$-intercept $c=-0.26\pm0.13$.  The errors were determined using a bootstrap method with Monte-Carlo sampling ($N=10,000$ simulations).  The Pearson's correlation coefficient is $r=0.927$ and the two-point probability of a chance correlation is $p=2.0\times10^{-10}$.  The value of $r$ is well in excess of the critical value for 22 dof at the $p=0.001$ level.  It is seen that the $k$ and $s$ values in $R$ are marginally consistent with the best-fitting line to within $1\sigma$, while those in $B$ and $V$ are entirely consistent with the relationship.   As discussed in \citet{Cano14}, this statistically significant $k$-$s$ (i.e. luminosity$-$stretch) relationship indicates that GRB-SNe have the potential to be used as standardizable candles \citep{CJG14,LiHjorth2014} in SN-cosmology.

\section{SN 2016jca - Bolometric Properties}
\label{sec:bolo}

We constructed a \emph{quasi}-bolometric LC from our optical observations (de-reddened, host- and AG-subtracted) in $griz$.  We followed the procedure outlined in detail in \citet{CMS14}, which briefly, includes creating SEDs for each epoch, fitting a linear spline to the data, and integrating the spline between the frequency limits of the reddest and bluest filters (i.e. assuming no flux beyond these limits).  We used the effective wavelengths given in \citet{Fukugita95}.  Then, the flux bolometric LC was transformed to a luminosity bolometric LC using a distance of 700~Mpc.  We estimated the luminosity errors by taking the average fractional uncertainty in each $griz$ SED (which includes the uncertainties in the photometry, AG model, and host photometry, all added in quadrature and propagated through the analysis), and applied this to the bolometric luminosity error. The final bolometric LC is shown in Fig.~\ref{FigArnett}.  The peak $griz$ luminosity is $L = 4.6\times10^{42}$~erg~s$^{-1}$, which occurs around +10.7~days (rest-frame).

\subsection{Temperature evolution}
\label{sec:temp_evo}

Under the assumption that SN~2016jca emits as a pure BB, and does not suffer any dilution effects (which is likely to be an over-simplification of reality, e.g. \citealt{Dessart05,Dessart15}), we fit the $griz$ SEDs with a Planck function to determine the BB colour temperature ($T_{\rm BB}$, in filters $griz$) and the radius ($R_{\rm BB}$) of the BB emitter.  Their evolution is plotted in Fig. \ref{Figtemp}.

%While $griz$ SEDs exist as early as $t-t_0=2.5$~d (rest-frame), the shape of the SED does not resemble that of a BB emitter, but is much steeper, owing to an additional component at bluer wavelengths, namely the $g$-filter (see Sect. \ref{sec:gbandLC}).  As such, we were unable to successfully fit the Planck function to the SEDs at times $<t-t_0=5.6$~d and hence derive colour temperatures and BB radii.

From the figure, the BB temperature has an initial value of $T_{\rm BB}~=~37\,000$~K at $t-t_0=5.6$~d, which then decreases rapidly, reaching a plateau of $T_{\rm BB} \approx 5000-6000$~K after approximately 20~days.  The BB radius is $\approx 3\times10^{14}$~cm during the first epoch, and reaches a maximum radius of $R_{\rm BB} \approx 3-4\times10^{15}$~cm around $t-t_0=20-30$~d, before decreasing to $R_{\rm BB} \approx 1-2\times10^{15}$~cm after 40 days.

\subsection{The Radioactive Heating Model}

Currently, the accepted physical processes that are thought to power GRB-SNe are heating arising from the interaction of $\gamma$-ray photons emitted during the decay process of nickel and cobalt into their daughter products (i.e. the radioactive heating model, \citealt{Arnett1982}), and energy input from a magnetar central engine, whose presence has been invoked for SLSNe-I and the very luminous GRB-SN~2011kl \citep{Greiner15}.  The first model is considered in this section, while the latter in the following subsection.

Two versions of the radioactive heating model (see Appendix \ref{sec:app_radioactive}) were fit to the $griz$ bolometric LC of SN~2016jca: one that assumes that all emitted $\gamma$-rays are thermalized in the SN ejecta (red model in Fig. \ref{FigArnett}), and another that allows a fraction of the $\gamma$-rays to escape into space without interacting with, or depositing energy into the SN (black model in Fig. \ref{FigArnett}).  In the latter model, an additional free-parameter is the $\gamma$-ray opacity ($\kappa_{\gamma}$).  Both models assume a grey optical opacity of $\kappa = 0.07$~cm$^2$~g$^{-1}$, and a peak photospheric velocity, as inferred from the Fe \textsc{ii} $\lambda$5169 line velocities, of $v_{\rm ph} = 29,700\pm1,500$~km~s$^{-1}$.

First, when the model that assumed full trapping of all emitted $\gamma$-rays (red line) was fit to the data, we find a nickel mass of $M_{\rm Ni} = 0.22\pm0.08$~M$_{\odot}$, an ejecta mass of $M_{\rm ej} = 5.8\pm0.3$~M$_{\odot}$, and a kinetic energy of $E_{\rm K} = 5.1\pm0.8\times10^{52}$~erg.

Next, we fit the partial-trapping model (black) to the bolometric LC, finding a nickel mass of $M_{\rm Ni} = 0.22\pm0.08$~M$_{\odot}$, an ejecta mass of $M_{\rm ej} = 5.9\pm0.3$~M$_{\odot}$, and a kinetic energy of $E_{\rm K} = 5.2\pm0.8\times10^{52}$~erg.  The bolometric properties obtained from both models agree very well, where the only difference is the slightly increased ejecta mass constrained by the partial-trapping model. We also found a $\gamma$-ray opacity of $\kappa_{\gamma} = 0.034$~cm$^2$~g$^{-1}$.  To put the value of the $\gamma$-opacity into context, \citet{Wheeler15} determined this value for a sample of $N=20$ SNe Ibc, finding $\kappa_{\gamma} = 0.010$~cm$^2$~g$^{-1}$ for the one GRB-SN in their sample (SN~1998bw), which is roughly three times smaller than that found here for SN~2016jca.  Overall, \citet{Wheeler15} find a range of $\gamma$-ray opacities $0.001 \leq \kappa_{\gamma} \leq 0.047$, for which SN~2016jca falls within the upper end of this range.

 %Visually, the latter model provides a better fit to the data.  

\begin{figure}
   %\centering
\includegraphics[width=\columnwidth]{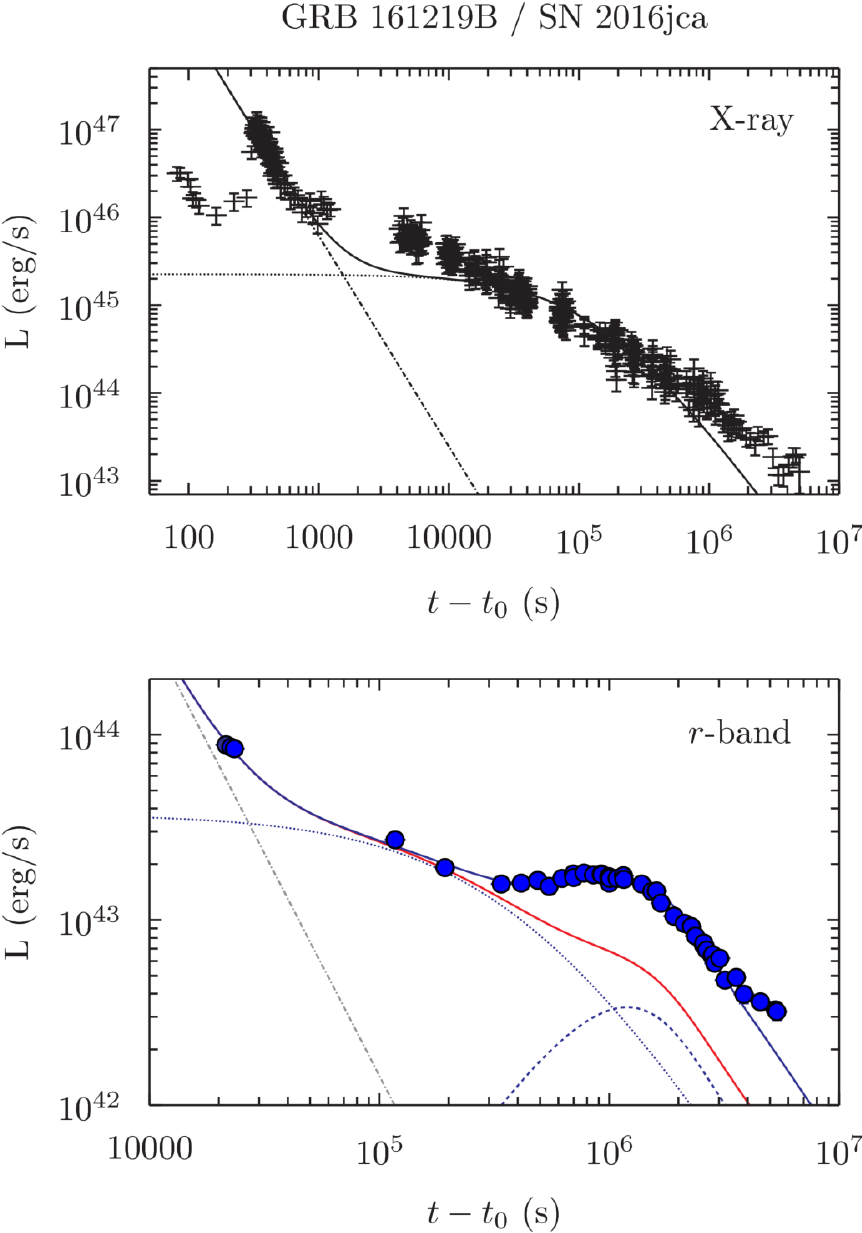}
      \caption{Magnetar model from \citet{CJM16} fit to the X-ray (\textit{Top}) and $r$-band (\textit{Bottom}) luminosity LCs. Times are shown in the rest-frame.   \textit{Top}: The sum of the SPL (dot-dashed line) and magnetar-powered AG (dotted line) phases is shown as the solid black line. Visual inspection shows that the model fails to reproduce the decay seen in the observations at $t-t_0>3\times10^5$~s.  The best-fitting model gives $B_{\rm 0} = 1.0\times 10^{15}$~G, and $P_{\rm 0} = 8.1$~ms.  \textit{Bottom}: The sum of the SPL (dot-dashed line), magnetar-powered AG (dotted line) and magnetar-powered SN (dashed line) phases is shown as the solid red line.  The solid blue line includes an additional multiplicative factor ($\Psi$) to get the magnetar-powered SN to match the observations.  The best-fitting model gives $B_{\rm 0} = 2.4\times 10^{15}$~G, and $P_{\rm 0} = 35.2$~ms, which are clearly discrepant with those determined from the X-ray.  In addition, we found a diffusion time scale of $t_{\rm diff} = 9.43\pm0.10$~days, and $\Psi = 4.2\pm0.1$.  The conclusion is that the magnetar model cannot self-consistently explain the X-ray and optical observations of GRB~161219B/SN~2016jca.  Even if SN~2016jca was powered in part by a magnetar central engine, an additional source of energy is needed to reproduce the observations, most likely a reservoir of radioactive nickel.  The required nickel mass to get the model to match observations is M$_{\rm Ni} \sim 0.4$~M$_{\odot}$.}
         \label{FigMagModel}
\end{figure}

For SN~2016jca, \citet{Ashall17} found a nickel mass of $M_{\rm Ni} = 0.4$~M$_{\odot}$, an ejecta mass of $M_{\rm ej} = 8$~M$_{\odot}$, and a kinetic energy of $E_{\rm K} = 5.4\times10^{52}$~erg.  These results were obtained from modelling both a bolometric LC (obtained over a wavelength range of $3000-10\,000$~\AA) and radiative transfer modelling of their spectral time series.  The results of each study are loosely consistent, though both the ejecta mass and nickel content therein found here are smaller than those found by \citet{Ashall17}.  It has been shown in previous studies \citep{Modjaz09,Cano2011b,Lyman14}, that including data over different frequency ranges affects the constructed bolometric LC differently.  Including bluer data causes the bolometric LC to peak earlier, while including NIR observation causes the LC to become wider and peak later.  Importantly, the inclusion of more data clearly makes the bolometric LC brighter and more luminous.  The wavelength range investigated here is $4900-9200$~\AA, which is smaller than that in \citet{Ashall17}, is clearly responsible for us finding smaller nickel and ejecta masses.  Finally, given the similar peak photospheric velocities considered in each paper ($v_{\rm ph} \approx 26\,000$~km~s$^{-1}$, their fig. 5), it is expected that we find similar explosion energies.

When visually comparing the bolometric LC of SN~2016jca to other GRB-SNe, it is seen that in relative terms, SN~2016jca is the least luminous.  One caveat however is that while an attempt has been made to compare bolometric LCs of the GRB-SNe over similar wavelength ranges, this is not always possible.  For example, the bolometric LC SN~2003dh was constructed from $UBVR$ observations \citep{Deng05}, SN~1998bw is from $BVRI$ \citep{Patat01}, SN~2006aj is from $BVRI$ \citep{Pian06}, SN~2010bh is from $griz$ \citep{Olivares12} and SN~2012bz is from $griz$ \citep{Schulze14}.  Next, relative to the "average" GRB-SNe \citep{Cano2016}, which has $M_{\rm Ni} = 0.4$~M$_{\odot}$ ($\sigma=0.2$~M$_{\odot}$), an ejecta mass of $M_{\rm ej} = 6$~M$_{\odot}$ ($\sigma=4$~M$_{\odot}$), and a kinetic energy of $E_{\rm K} = 2.5\times10^{52}$~erg ($\sigma=1.8\times10^{52}$~erg), SN~2016jca synthesized less radioactive material, but a "typical" mass of ejecta.  SN~2016jca is more energetic than the average GRB-SNe because its peak photospheric velocity is more rapid than that of the average GRB-SN ($v_{\rm ph} = 20\,000$~km~s$^{-1}$ ($\sigma=8000$~km~s$^{-1}$) by more than $1\sigma$.  As noted in other works \citep{Mazzali14,Ashall17}, the kinetic energies and ejecta masses determined from 1D analytical modelling should be considered as upper limits to their true values as they do not consider the true aspherical nature of the ejecta (e.g. \citealt{Mazzali01,Maeda02,Maeda06,WangWheeler08}).

\subsection{The Magnetar Model}

Next, we fit both the optical and X-ray data to see if the luminosity could plausibly be explained within the context of the magnetar model (see Appendix \ref{sec:app_magnetar}), using the model from \citet{CJM16}.  For the magnetar model to be deemed viable, the initial spin period ($P_{\rm 0}$) and magnetic field strength ($B_{\rm 0}$) of the magnetar central engine should be consistent when fitting the X-ray and optical data independently, otherwise the model is rejected.

\begin{figure*}
   \centering  % trim={<left> <lower> <right> <upper>}
   \includegraphics[width=\hsize,  trim={0 0 0 0}]{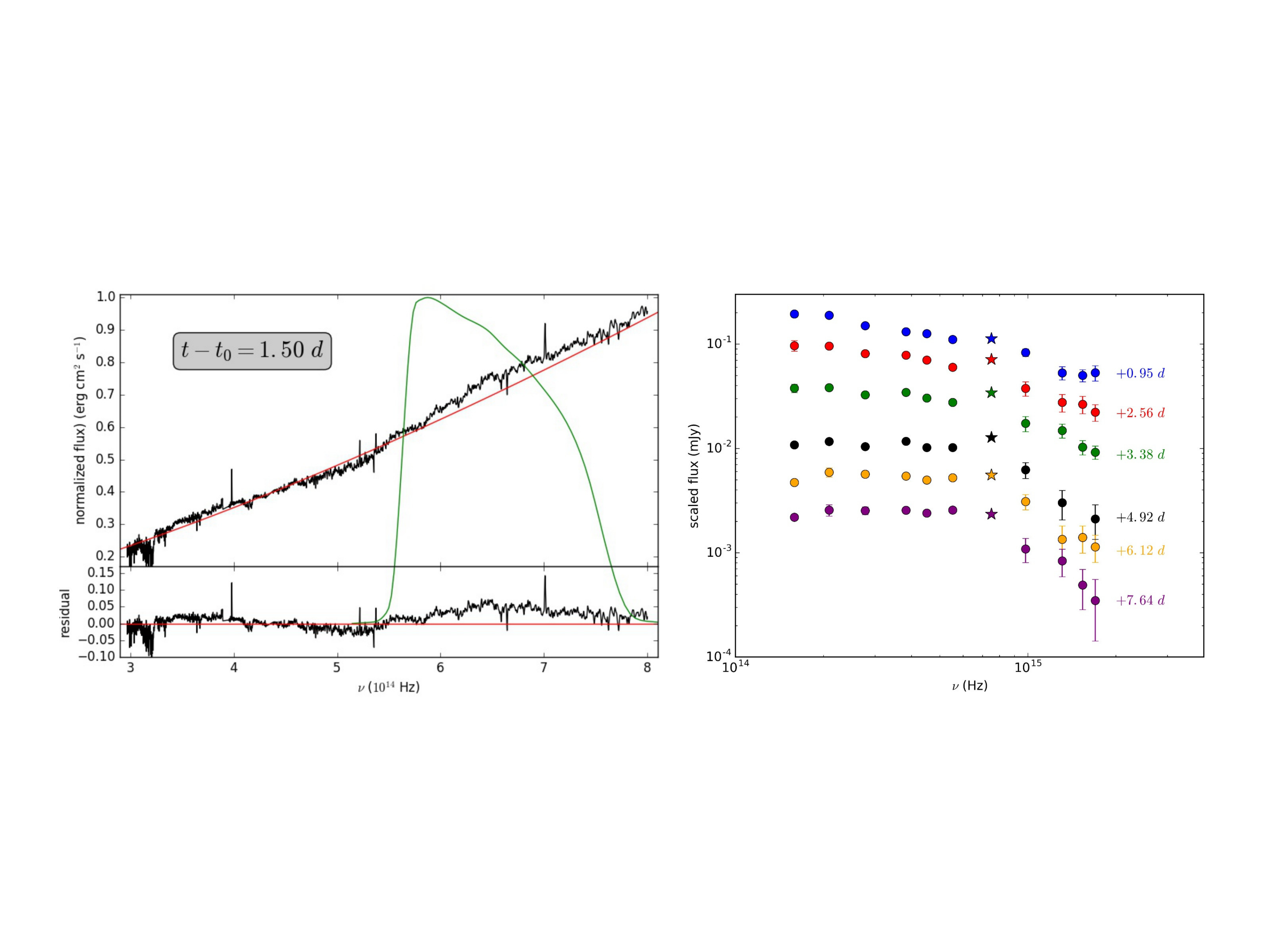}
      \caption{Inspection of the $g$-band excess. \textit{Left}: XS spectrum obtained at $t-t_0=1.5$~day (observer-frame).  A SPL (red line) was fit to the spectrum in the wavelength range $6000 \le \lambda \le 9000$~\AA~ ($3-5\times10^{14}$~Hz).  Between $3000\sim5500$~\AA~($5.5-8\times10^{14}$~Hz), excess above the best-fitting SPL is seen.  Over-plotted is the $g$-band transmission curve, which shows that the excess occurs only in this filter.  \textit{Right}: Optical \& UV SED evolution (AG+SN+host), using data taken from our ground-based telescopes and \emph{Swift}-UVOT.  The $g$-band photometry are shown as stars.  At early times ($t-t_0=0.95$~day), little excess is seen in the $g$-band.  However, between $t-t_0 = 2.56-6.12$~days, a clear excess is seen, which disappears by 7.64~days.  Intriguingly, excess is not seen in the bluer UVOT filters. }
   \label{FiggSED}
\end{figure*}

First, we fit the \emph{Swift}-XRT (top panel in Fig. \ref{FigMagModel}).  When fitting the SPL and magnetar-powered AG phases, the free parameters are the normalization constant ($\Lambda$) of the former, and the luminosity ($L_{0}$) and duration ($T_{o}$) of the latter phases.  The SPL index was fixed to $\alpha=\Gamma_{\gamma} + 1 = 2.4$.  The best-fitting model gives $L_{0} =$~($2.25\pm0.08$)~$\times 10^{45}$~erg~s$^{-1}$ and $T_{0} =$~($1.39\pm0.06$)~$\times 10^{5}$~s.  In turn this implies $B_{\rm 0} = 1.0 \times 10^{15}$~G, and $P_{\rm 0} = 8.1$~ms.  Visual inspection of the LC (top panel in Fig. \ref{FigMagModel}) shows that the model is a poor fit to the data, where the observations after $t-t_0 = 3\times 10^5$~s (rest-frame) fade at a slower rate than that of the magnetar model ($t^{-2}$). 

Next, we fit the magnetar model to an $r$-band luminosity LC (observer-frame $griz$) of GRB~161219B/SN~2016jca (bottom panel of Fig. \ref{FigMagModel}), using an identical approach as that used in \citet{CJM16}.  The free parameters in this model, in addition to $L_0$ and $T_0$, are the diffusion timescale ($t_{\rm diff}$) and a multiplicative factor ($\Psi$), which is needed in cases where the luminosity of the magnetar-powered SN, as determined from the magnetar-powered AG phase, is under-luminous relative to the observations.  The best-fitting model gives $L_{0} =$~($3.73\pm0.26$)~$\times 10^{43}$~erg~s$^{-1}$, $T_{0} =$~($4.43\pm0.51$)~$\times 10^{5}$~s, $t_{\rm diff} = 9.43\pm0.10$~days, and $\Psi=4.2\pm0.1$. In turn this implies $B_{\rm 0} = 2.4 \times 10^{15}$~G, and $P_{\rm 0} = 35.2$~ms.  As before for the X-ray LC, the late-time decay of the observations clearly deviates from that of the magnetar model.

The clear mismatch between the values determined from fitting the X-ray and optical data independently imply that the magnetar model cannot satisfactorily, and self-consistently, describe all phases of GRB~161219B/SN~2016jca.  In the unlikely scenario that SN~2016jca is powered in part by emission arising from a magnetar central engine, an additional source of energy is needed to explain its $r$-band luminosity, which is likely to be radioactive nickel.  In this case, an additional mass of M$_{\rm Ni} \sim 0.4$~M$_{\odot}$ is required. (Note that the required nickel mass in the magnetar model is larger than that inferred from modelling of the bolometric LC with the radioactive-heating model.  In the magnetar model, we are fitting the AG and SN simultaneously, and the AG decays as $t^{-2}$.  When modelling the AG in Sect. \ref{sec:AG}, the AG is seen to decay as $t^{-0.8}$, thus the AG decays slower and contributes more flux to the later SN phase (which is then removed) than in the magnetar model. Hence, the bolometric LC constructed from the AG-subtracted data is fainter, and requires less nickel to explain its luminosity.)

We note that if the Fe \textsc{ii} $\lambda$5169 line velocities are a suitable proxy for the photospheric velocity, their evolution is inconsistent with that of the 1D magnetar model (e.g. \citealt{KasenBild10,Wang17}; see \citealt{Cano2016} for further discussion), which predicts a flat evolution.  Inspection of Fig. \ref{FigLineVel} instead reveals a steady decline in line velocity.  Moreover, the maximum kinetic energy determined via the radioactive heating model is in excess of that expected from the magnetar model (e.g. \citealt{Usov92}), where it is suggested that only up to $\sim 2\times10^{52}$~erg energy is available to power the SN (though see \citealt{Metzger15}).  That we find more than twice this value adds additional credence to the notion that the compact object formed during the core-collapse of the progenitor of SN~2016jca was a black hole, rather than a neutron star.  This conclusion is contrary to the results of \citet{Ashall17}, who, despite finding a kinetic energy in excess $5\times10^{52}$~erg, suggest the compact object formed at the time of core-collapse was a magnetar, which is based primarily on the wide jet angle inferred from modelling their optical and X-ray LCs.

\section{The Mystery of the Pre-Maximum \textit{g}-band Bump}
\label{sec:gbandLC}

As seen in Fig. \ref{FigLC_optical}, the SN appears to be double peaked in the $g$-band, though intriguingly such behaviour is not observed for the other optical filters ($riz$).  Early peaks (i.e. those occurring before the main peak) have been observed for other GRB-SNe, namely SN~2006aj \citep{Campana06} and SN~2010bh \citep{Cano2011b,Olivares12}. Pre-maximum bumps have also been observed for type Ic SNe not associated with a GRB, including SLSNe-Ic \citep{Leloudas12,Nicholl15,Smith16} and type Ic SN iPTF15dtg \citep{Taddia16}.  Detecting a pre-peak bump requires daily cadence, especially during the first 10 days, and a smoothly evolving AG (i.e. unlike that observed for SN~2003dh, \citealt{Matheson03}).  This rules out the possibility of detecting such a bump for many GRB-SNe where the observational cadence was insufficient during these times.  For those GRB-SNe that were well observed, no such bump was detected (e.g. SN~1998bw, SN~2012bz, SN~2013dx, GRB~140606B).  Interestingly, a slight excess of flux was found in the observer-frame $r$-band filter of SN~2013fu, associated with GRB~130215A ($z=0.479$, corresponding to rest-frame $\lambda = 6290/(1+z)=4253$~\AA), which is defined by only three data-points, and the excess was not discussed by the authors \citep{Cano15}.

The origin of the pre-maximum flux for SN~2006aj has been debated by several authors.  \citet{Campana06} explained the achromatic pre-maximum peaks as cooling shock-heated material (from the initial shock breakout; SBO).  A thermal component was also seen in the X-ray, which cooled and moved into the UV and optical regimes. In this model the observed features arose from the breakout of a shock driven by a mildly relativistic shell into a dense, and optically thick, stellar wind.  An alternative model to explain the achromatic early peaks was proposed by \citet{Margutti15} and \citet{Nakar15}, where the breakout of the thin shell is from an extended (a few hundred solar radii) low-mass (a few hundredths of a solar mass) envelope surrounding the exploding star.

For the non-GRB type Ic SN iPTF15dtg, \citet{Taddia16} considered several models to explain the pre-maximum peak, including a SBO cooling tail (e.g. \citealt{PiroNakar13}), a magnetar-driven SBO tail \citep{Kasen16}, the extended-envelope scenario of \citet{Nakar15}, and SN ejecta interacting with a companion star in a binary system \citep{Kasen10}.  In the cases of type Ic SN iPTF15dtg and SLSNe-Ic LSQ14bdq \citep{Nicholl15} \& DES14X3TAZ \citep{Smith16}, the extended-envelope scenario provided the most realistic explanation of the early achromatic peaks.

For SN2016jca, there is one key difference with respect to the aforementioned SNe: the early peak is \emph{not achromatic}, but only appears in the $g$-band.  Already this rules out all of the aforementioned scenarios, all of which predict pre-maximum peaks in several filters, and not just one.

So how can this early peak be explained?  The first point of interest is determining if it is even real $-$ it could instead be a relic of an improper data-reduction and calibration method.  In order to determine whether the early $g$-band peak is real, we inspected our spectra to look for evidence of any excess in the wavelength range corresponding to $g$-band, and during the same time window.  The earliest spectrum presented here is the XS spectrum obtained at at $t-t_0=1.5$~day (observer-frame).  To check for excess, we fit the entire spectrum with a SPL, (Fig.~\ref{FiggSED}), where the $g$-band transmission curve is over-plotted for reference.  Between $3000\sim5500$~\AA~($5.5-8\times10^{14}$~Hz), excess above the best-fitting SPL is clearly seen.

Thus, two different telescopes/instruments confirm that the early $g$-band excess is seen.  But what about bluer wavelengths; is excess also observed?  To address this, we compiled the \emph{Swift}-UVOT observations obtained of SN~2016jca up to $t-t_0=8$~days.  We then investigated several NIR to UV SEDs to check for excess at other wavelengths (right panel of Fig. \ref{FiggSED}).  Six epochs are shown for $t-t_0 = 0.95, 2.56, 3.38, 4.92, 6.12$ and $7.64$~days.  The epoch at $t-t_0 = 0.95$~days offers little evidence for $g$-band excess, though the excess is clearly seen between $t-t_0 = 2.56-6.12$~days, and then disappears by $7.64$~days.  Moreover, visual inspection of the SEDs reveals that \emph{only} the $g$-band displays evidence of flux excess.

The cause of this $g$-band excess is not immediately obvious.  It is unlikely to be related to one or more emission lines as none are observed in this wavelength region in the spectral time-series shown in Fig. \ref{FigSpectra} (though we note that \citealt{Ashall17}, some excess of flux is also seen in their spectrum at $t-t_0=3.73$~days around $\lambda=4000$~\AA).  Moreover, it evolves quite rapidly: it is not convincingly seen in the SED at $t-t_0=0.95$~day (0.83~days rest-frame), and has disappeared by 7.64~days (6.6~days rest-frame).  This chromatic behaviour is not readily explained by the aforementioned theoretical models, which ultimately leaves its physical origin an unsolved mystery.

\begin{figure}
   %\centering
   \includegraphics[width=\hsize]{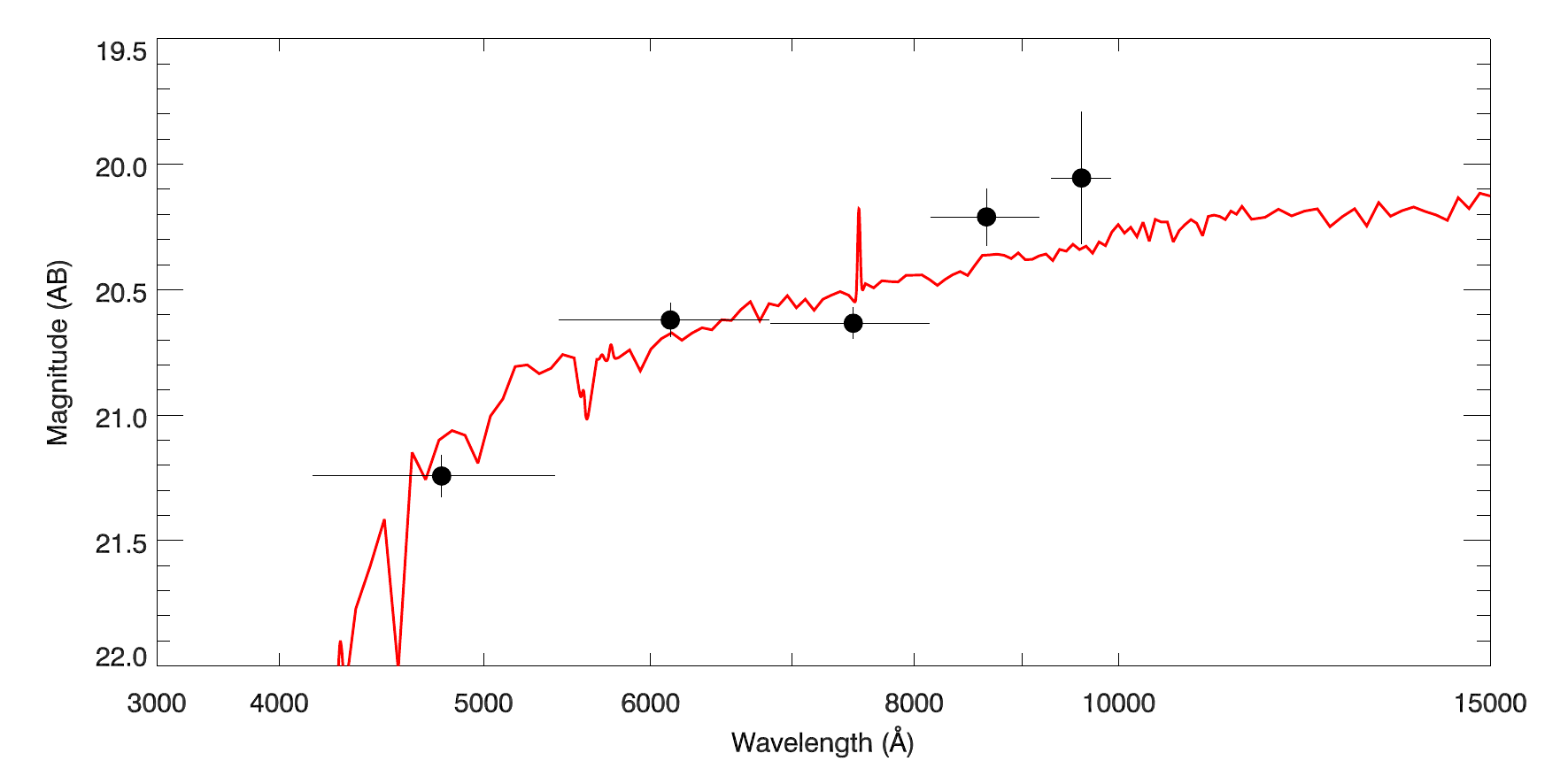}
      \caption{Optical ($grizY$) photometry (for a $4\farcs0$ circular aperture) of the host galaxy of GRB~161219B.  We fit the models of \citet{Bruzual03} to the optical SED, finding best-fitting parameters of: an age of $(0.90_{-0.16}^{+5.98})\times10^9$~years, a stellar mass of log(M$_{*}/$M$_{\odot})=(8.88_{-0.10}^{+1.03})$, a SFR of $0.25_{-0.17}^{+0.30}$~M$_{\odot}$~yr$^{-1}$ and negligible intrinsic extinction.}
         \label{FigHostSED}
\end{figure}

\begin{figure}
   %\centering
   \includegraphics[width=\columnwidth]{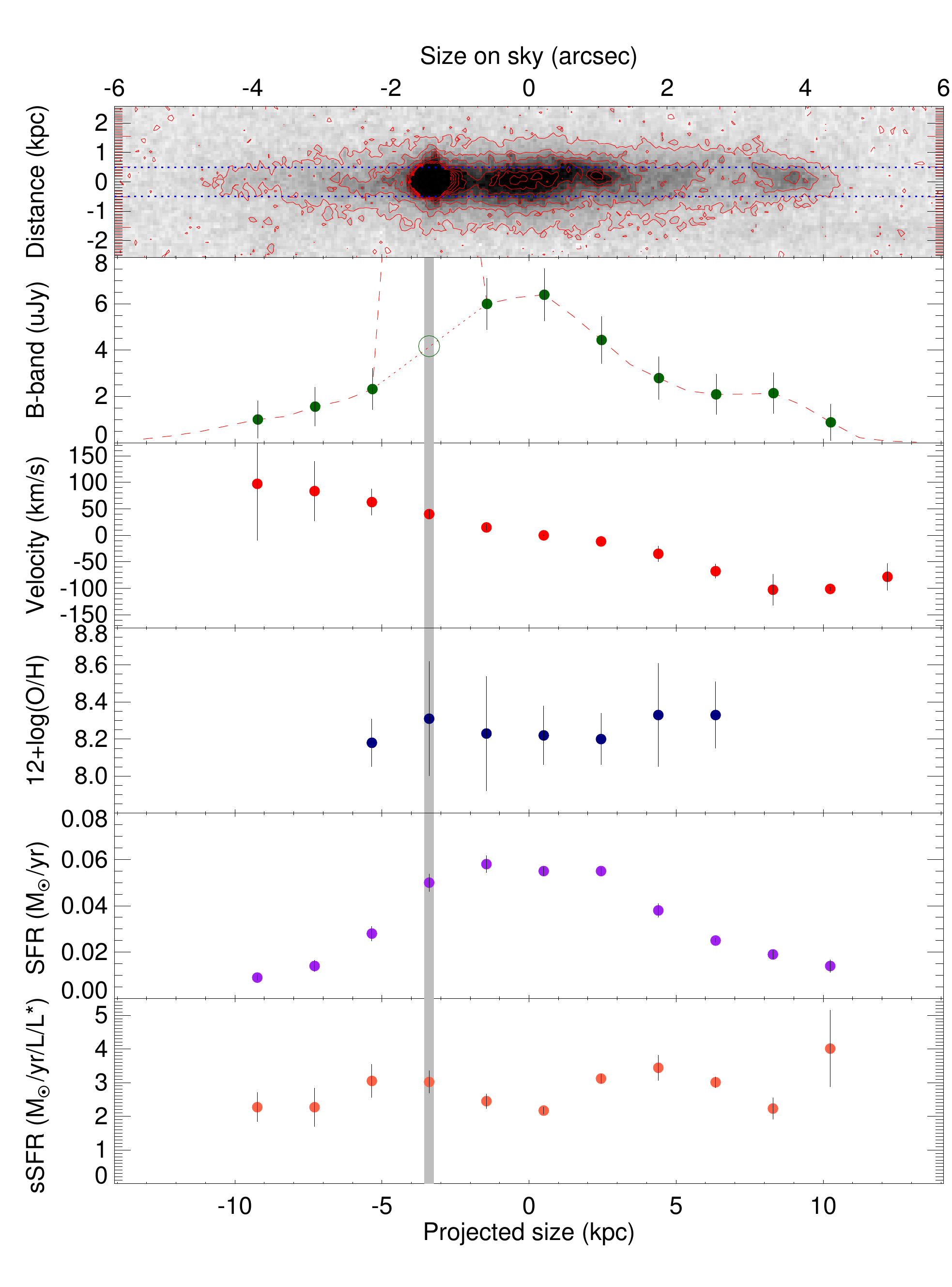}
      \caption{Host galaxy properties along the slit on the GTC observation of the 26 January 2017, as compared to the \emph{HST} image. The first panel shows one of the \emph{HST} images with contours superposed in red and the position and width of the GTC slit indicated by the dotted blue lines. The second panel is the $B$-band flux profile  along the slit using the flux obtained from the 0$^{\rm th}$ order L200LP grism from \emph{HST} as reference. The third panel shows the relative velocity measured using the [O \textsc{ii}], [O \textsc{iii}] and H$\alpha$ lines. The fourth panel shows the metallicity using the N2 parameter \citep{Marino13}. The fifth panel shows the SFR derived from the H$\alpha$ emission. The last panel is the SFR weighted by the $B$-band magnitude in panel 2. The vertical grey line marks the location of SN~2016jca. }
         \label{FigHostProp}
   \end{figure}

\section{The Host Galaxy}
\label{sec:host}

\begin{figure}
   %\centering
   \includegraphics[width=\columnwidth]{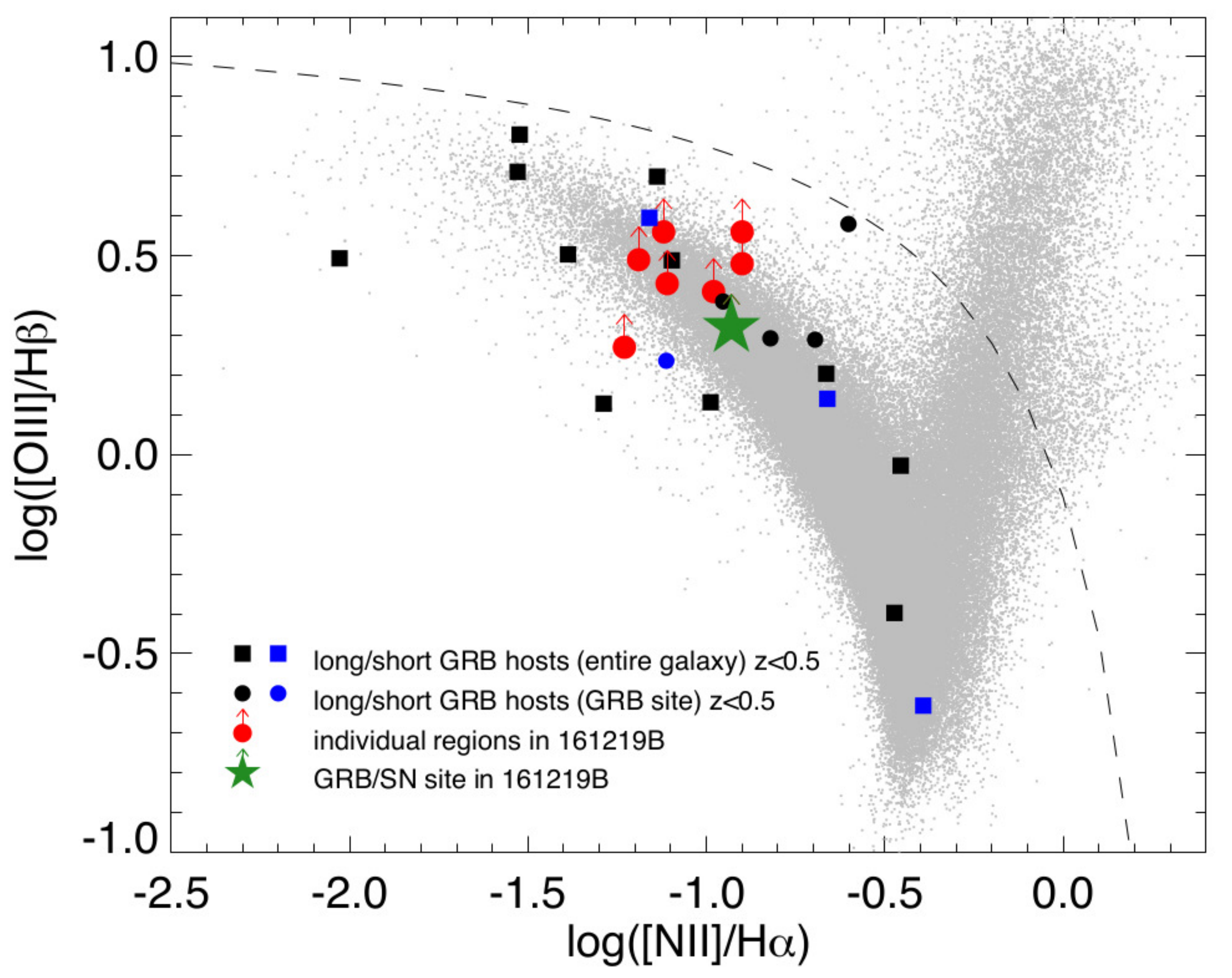}
      \caption{BPT diagram showing the different regions in the host of GRB~161219B, where the GRB site is highlighted by the green star. Values presented here are only lower limits as H$\beta$ could not be measured due to its location on top of an atmospheric emission line. Grey dots are galaxies from SDSS (DR9), while the squares and dots are other GRB hosts/GRB sites (the latter for cases of resolved galaxies) at z$<$0.5. The dashed line marks the dividing line between SF regions and those dominated by AGN activity. Data are taken from \cite{Christensen08, Berger09, Han10, Levesque10, Perley12, Thoene14, deUgarte14b, Schulze14, Stanway15,Kruehler15}.}
         \label{FigHostBPT}
   \end{figure}

The field of GRB~161219B was observed by Pan-STARRS1 in $grizY$ prior to explosion. These images show a host galaxy that is morphologically consistent with an edge-on spiral. The GRB appears to be located close to the disk plane, at a distance of $1\farcs5\pm0\farcs2$ from the galaxy bulge, equivalent to a projected distance $3.9\pm0.5$ kpc at a redshift of $z=0.1475$.  A recent study by \citet{Lyman17} found the median offset of a sample of $N=39$ GRBs from their apparent host centres of $1.0\pm0.2$~kpc, which implies that GRB~161219B occurred at a relatively further distance from its host's centre than most GRBs.

Photometry of the host was performed on these images using a circular aperture with a radius of $4\farcs0$ that encircled the complete galaxy light visible in the Pan-STARRS1 images, and were calibrated using Pan-STARRS1 DR1 (PS1) field stars.  We found AB magnitudes of: $g = 21.24\pm0.08$, $r = 20.62\pm0.07$, $i = 20.63\pm0.08$, $z = 20.21\pm0.12$, $Y = 20.06\pm0.26$, which are corrected for foreground extinction.  Note that these magnitudes differ from those in Table \ref{table:photometry}, which are for a smaller aperture of $2\farcs2$.

Next, a set of galaxy templates were fit to the derived host magnitudes using \texttt{LePhare} (version 2.2, \citealt{Arnouts99,Ilbert06}). The templates were based on the models from \citet{Bruzual03}. The photometry of the host galaxy of GRB~161219B is best fit by a galaxy template (see Fig.~\ref{FigHostSED}) with a stellar mass of log(M$_{*}/$M$_{\odot})=(8.88_{-0.10}^{+1.03})$, a star-formation rate (SFR) of $0.25_{-0.17}^{+0.30}$~M$_{\odot}$~yr$^{-1}$, an age of $(0.90_{-0.16}^{+5.98})\times10^9$~years, and a negligible intrinsic extinction. We note that the constraints derived from the host galaxy photometric fit are not overly constraining due to the limited wavelength coverage.

\emph{HST} imaging obtained with WFC3 (proposal \#14901, PI: A. Levan) shows the host galaxy in much more detail, which again resembles an edge-on spiral (see Fig.~\ref{FigFC}) with an elongated disk that extends $8\farcs5\times0\farcs8$ ($22\times2$~kpc), a central bulge, and at least two distinctive knots on the disk, possibly due to star-forming (SF) regions. In the \emph{HST} image we measure a distance of SN~2016jca of $1\farcs30\pm0\farcs05$ ($3.38\pm0.13$~kpc) to the centre of the galaxy (NB: defining the precise location of galactic core poses the largest source of uncertainty in this calculation).

The spectra of SN~2016jca obtained on the 22-January-2017 has the slit positioned along the edge-on galaxy (see Table \ref{table:spectra_obs_log}). We extracted the 1D spectra in fixed-width bins ($1\farcs5$) in steps of $0\farcs75$ and analyzed them separately. The flux values and properties in the different regions are listed in Table \ref{table:host_emlines}, while the properties along the slit are plotted in Fig. \ref{FigHostProp}. We also calculated the specific SFR (SSFR) weighted by the (rest-frame) $B$-band magnitude (see Table \ref{table:host_emlines}). To derive these values, we obtained a magnitude from the flux of the spectra in the redshifted range of a Johnson $B$-band filter and then used these values to scale the flux obtained from the L200LP grism in the 0$^{\rm th}$ order in the \emph{HST} slit-less spectra, whose centre is around 5000~\AA, and which give us a better spatial resolution than the ground-based data. To obtain the flux at the SN position where the continuum is dominated by the SN emission, we extrapolated the value from the neighbouring regions. 

There is little variation across the galaxy, which is not surprising given its edge-on viewing angle, hence the spectra are all dominated by the light from the outer spiral arms. Neither the star-formation rate (SFR) nor the metallicity are extreme/peculiar at the SN position compared to the rest of the host (see Table \ref{table:host_emlines}). Indeed, the SFR and sSFR are highest in the SF region at the opposite side of the galaxy. The integrated host spectrum has a metallicity of 0.4 solar (12+log(O/H)=8.28) and a SFR of 0.17~M$_{\odot}$~yr$^{-1}$, consistent with the value obtained from the SED fitting of the host. The sSFR of the entire galaxy is 2.91~M$_{\odot}$~yr$^{-1}$L~L$_{*}^{-1}$, and we find a mass-weighted SFR of 0.18~Gyr$^{-1}$.

We find that the mass and the sSFR are consistent with the mean value of GRB hosts at $z<0.5$ (e.g. \citealt{Perley16,Schulze16}), but the SFR is smaller than the average. \cite{Kruehler15} find that the SFR increases with redshift, which is partially an effect of increasing host mass with redshift since SFR and stellar mass are known to be correlated (with exceptions), i.e. the so-called SFR-main sequence \cite[e.g.][for low and high redshifts respectively]{Elbaz07,Bouwens12}. The stellar mass of GRB~161219B's host is somewhat above-average for hosts at a similar redshift \cite[see e.g.][]{Kruehler15,Perley16,VerganiBAT6}, so one might expect a higher SFR that measured here; instead we find the opposite, i.e. that the galaxy has one of the lowest SFRs measured for any GRB host (see \citealt{Kruehler15}). Edge-on galaxies often show lower measured SFRs as part of the light is hidden behind dust lanes, however, this does not seem to be an issue here as we measure very low extinction in the SED fit. 

The metallicity of the host is rather typical for a long-duration GRB host \citep{Kruehler15}, which appears to support the notion of previous results that GRB hosts do not show an extremely low metallicity preference, as required by most current GRB progenitor models. In Fig. \ref{FigHostBPT} we plot the line ratios from the different parts of the galaxy into the Baldwin-Phillips-Terlevich (BPT) diagram, which allows us to distinguish between SF- and AGN-driven regions, and to some extent the age and metallicity of each region, depending on the ionization parameter. In general, younger and more metal-poor galaxies are found towards the upper left of the BPT diagram. All regions within the host of GRB~161219B occupy very similar regions in the diagram, and they are well within the part of the diagram typically occupied by GRB hosts at low redshifts, but somewhat more extreme than the bulk of SF galaxies found in the SDSS.

In summary, the host is a rather typical GRB host at its redshift, with the only difference of a relatively lower SFR and sSFR. Most GRB hosts at low $z$ seem to be dwarf galaxies, but there is a growing fraction of GRBs occurring in spiral galaxies (e.g. GRB~980425 \citealt{Fynbo00,Christensen08,Kruehler17}, GRB~060505 \citealt{Thoene14}\footnote{Although there is debate regarding the long/short nature of this GRB, e.g. \citet{Ofek07}}, GRB~111005A \citealt{Michalowski16}), though all of them have small stellar masses. Interestingly, in two of those spiral hosts (GRBs~060505 and 111005A), no SNe associated with the GRBs were detected despite their very low redshift \citep{Fynbo06, Michalowski16}. However, they both were rather different galaxies: GRB~060505 has a low metallicity and high SFR, particularly at the GRB site \citep{Thoene08,Thoene14}, while the host of GRB~111005A has a super-solar metallicity and an even lower sSFR than GRB~161219B \citep{Michalowski16}.%, although it is clearly affected by dust extinction due to its high inclination.

\begin{table*}
\scriptsize
\centering
\setlength{\tabcolsep}{1pt}
\setlength{\extrarowheight}{2pt}
\caption{GRB 161219B host galaxy emission lines and properties}
\label{table:host_emlines}
\begin{tabular}{lccccccccccccc}
\hline																
region &	[O\textsc{ii}] 	& [O\textsc{iii}]			& [O\textsc{iii}]	&	H$\alpha$	& [N\textsc{ii}] 	& [S\textsc{ii}]	& [S\textsc{ii}] & SFR   & sSFR & 12+log(O/H)	& log([O\textsc{iii}]/[O\textsc{ii}]) & log([N\textsc{ii}]/H$\alpha$) & log([O\textsc{iii}]/H$\beta$) \\
	&				&$\lambda4958$	&		&				&			&$\lambda6717$		&$\lambda6732$	& (M$_{\odot}$~y$^{-1}$) & (M$_{\odot}$~y$^{-1}$~L$^{-1}$~L$_{*}^{-1}$)&		&			&		&		\\ \hline
SN-1.5&	0			&0				 &0			&	1.88$\pm$0.26&0			&0				 &	0		    	  & 0.009$\pm$0.001 & 2.27$\pm$0.44	&0				&0			&0		&0			\\
SN-1	&4.38$\pm$1.2		&0				&1.31$\pm$0.66&	3.11$\pm$0.55	&0			&0				&0				&0.014$\pm$0.002	& 2.27$\pm$0.57	&0				&--0.40		&0		&0.17		\\
SN-0.5&7.55$\pm$0.8	&0				&3.26$\pm$0.45&	6.14$\pm$0.71	&0.36$\pm$0.08&0.84$\pm$0.24	&1.76$\pm$0.29	&0.028$\pm$0.003	& 3.05$\pm$0.50	&8.18$\pm$0.13	&--0.24		&--1.23	&	0.27\\
SN&	12.2$\pm$1.9		&2.34$\pm$0.6		&6.4$\pm$1.09	&	10.73$\pm$0.84	&1.28$\pm$0.6	0 &1.59$\pm$0.34	&2.18$\pm$0.42	&0.050$\pm$0.004	& 3.02$\pm$0.33	&8.31$\pm$0.31	&--0.16		&--0.93	&	0.32\\
SN+0.5&15.9$\pm$2.1	&0				& 9.3$\pm$1.0	&	12.27$\pm$0.79	&0.97$\pm$0.60  &2.46$\pm$0.96	&2.19$\pm$0.569	&0.058$\pm$0.003	& 2.45$\pm$0.22	&8.23$\pm$0.31	&--0.11		&--1.11	&	0.43	\\
SN+1&18.7$\pm$2.1	&3.8$\pm$0.8		&12.4$\pm$0.9	&	11.95$\pm$0.43	&0.94$\pm$0.33 &2.32$\pm$0.28	&2.40$\pm$0.30	&0.055$\pm$0.002	& 2.19$\pm$0.11	&8.22$\pm$0.16	&--0.05		&--1.12	&0.56 \\	
SN+1.5&15.4$\pm$1.2	&3.43$\pm$0.5		&10.2$\pm$0.7	&	11.74$\pm$0.39	&0.79$\pm$0.2	6  &2.14$\pm$0.30	&2.25$\pm$0.42		&0.055$\pm$0.002	& 3.11$\pm$0.15	&8.20$\pm$0.14	&--0.06		&--1.19	&0.49 \\
SN+2&10.4$\pm$0.14	&2.06$\pm$0.4 	&7.03$\pm$0.6	&	8.19$\pm$0.65	&1.06$\pm$0.45&1.27$\pm$0.18	&3.09$\pm$0.55		&0.038$\pm$0.003	& 3.44$\pm$0.38	&8.33$\pm$0.28	&--0.05		&--0.90	&0.48 \\
SN+2.5&6.93$\pm$0.75	&2.10$\pm$0.3		&5.49$\pm$0.3	&	5.33$\pm$0.21	&0.67$\pm$0.20&0				&0				&0.025$\pm$0.001	& 3.01$\pm$0.16	&8.33$\pm$0.18	&--0.02		&--0.90	&0.56 \\
SN+3&4.47$\pm$0.7	&2.06$\pm$0.3		&4.33$\pm$0.76&	4.13$\pm$0.43	&0			  &0				&0				&0.019$\pm$0.002	& 2.22$\pm$0.33	&0				&0.1			&0		&0.56 \\
SN+3.5&0				&0				&0			&	3.07$\pm$0.62 &0			&0				&0				&0.014$\pm$0.002	& 4.01$\pm$1.14	&0				&0			&0		&0		\\	\hline
total host&45.9$\pm$3.7	&7.78$\pm$1.3		&26.0$\pm$2.2	&	35.9$\pm$1.5	& 3.80$\pm$1.05&5.56$\pm$1.5	&6.33$\pm$1.6		&0.17$\pm$0.01	& 2.91$\pm$0.24	&8.28$\pm$0.15	&--0.13		&--0.98	&0.41		\\
\hline	
\end{tabular}
%\medskip
\begin{flushleft}
The regions mark the centre of the extracted spectrum relative to the SN location in units of steps (one step corresponds to $1\farcs5$), hence the spectra are overlapping. Fluxes are in units of 10$^{-17}$~erg~cm$^{-2}$~s$^{-1}$. For the ratio [O\textsc{iii}]/H$\beta$, the flux of H$\beta$ is derived from H$\alpha$ assuming zero extinction and Case B recombination (H$\alpha$/H$\beta$ = 2.76). Errors in the SFRs only reflect the error in the H$\alpha$ flux, while the sSFR also considers the error in the $B$-band magnitude.\\
\end{flushleft}
\end{table*}

\section{Discussion \& Conclusions}
\label{sec:conclusions}

We presented optical and NIR photometry, and optical spectroscopy of GRB~161219B and its associated SN~2016jca.  The early optical/NIR AG is characterised by a shallow decline, with a temporal index of $\alpha=0.6-0.8$.  The shallower index results from the NIR LCs, which are contaminated with host flux.  Instead, the host-subtracted optical data in $griz$ have a steeper decay index of $\alpha\approx0.8$, which is precisely that found from modelling the X-ray LC.  We also modelled several early NIR to X-ray SEDs with a BPL, finding spectral indices of $\beta_{\rm opt} \approx 0.45$ and $\beta_{\rm X} \approx 0.95$, with a break frequency of $\nu_{\rm BB} \approx 1.75\times10^{15}$~Hz, i.e. in the mid-to-far UV range.

The optical and X-ray spectral indices are consistent with synchrotron emission arising from a fireball colliding with  circumburst material (e.g. \citealt{{Sari98}}).  In this scenario, electrons accelerated by the fireball are cooling slowly, with an electron index of $p=1.9$. In our modelling, we find that the cooling break lies between the optical and the X-rays, which is the same conclusion found by \citet{Ashall17}. The values of the spectral and temporal indices are consistent with a fireball colliding with a homogeneous medium: from the closure relations between $\alpha$ and $\beta$ (e.g. \citealt{Sari99,ChevLi00}), our measured values of $\beta$ suggest a temporal index in the optical of 0.68 (as compared to the measured values of 0.61-0.86) and in the X-rays of 0.93 (as compared to the measured value of 0.79) before the jet break, and of 1.90 after the jet break (compared to the measured value of 1.93). We see that, although there are mild inconsistencies in the temporal slopes (which could be due to a departure from a uniform medium, i.e. it is slightly clumpy) the consistency between model and data is very good. We also note that the data are inconsistent with a stellar wind density profile surrounding the GRB progenitor, as it would require temporal decays in the optical of 1.18, which are inconsistent with the measured value.

When fitting the NIR to X-ray SEDs, a strong thermal component was found at $t-t_0=0.26$~days ($T_{\rm BB} \approx 0.16\times10^{6}$~K), which contributed roughly 70\% of the total observed flux.  The strength of the BB component decreased to $<10$\% at +1.47~days, and disappeared completely by $+2.55$~days.  The radius of the thermal component was found to be $R_{\rm BB} \approx 6\times10^{14}~$cm, which is much larger than those found for e.g. GRB~060218 and GRB~120422A by almost one order of magnitude.  Interestingly though, the NIR to X-ray SED of GRB~120422A at $t-t_0=0.267$~days had a similar temperature to that found for GRB~161219B at a similar post-explosion epoch ($T_{\rm BB} \approx 0.19\times10^{6}$~K), which was interpreted by \citet{Schulze14} as emission arising from a cooling, expanding stellar envelope after the passage of the shock breakout through it.  No pre-maximum bumps were observed for SN~2012bz, meaning a similar interpretation of the thermal component for GRB~161219B is appealing.

%In terms of the closure relations (e.g. \citealt{Racusin09}, table 1) between the temporal and spectral indices, it is possible to infer the presence of an ISM vs. wind-like medium local to the GRB progenitor, and possibly infer properties of the jet (spreading vs. non-spreading).  Before +1~day, (i.e. $1<p<2$) the indices are marginally consistent with an ISM-like medium and slow electron cooling, where the observed frequencies are between $\nu_{\rm m}$ and the colling frequency $\nu_{\rm c}$.  The observed values are inconsistent with a wind-like medium (both slow and fast cooling), and the jet properties cannot be constrained.  For times after 1.5~days ($\beta=1.3-1.4$, $p=2.8$), the CSM and jet properties cannot be constrained.

Next, we demonstrated that SN~2016jca was less luminous and evolved more rapidly than the comparison/template SN~1998bw.  Using the blueshifted line velocities of Fe \textsc{ii} $\lambda5169$ as a proxy for the photospheric velocity, we find that SN~2016jca has a peak photospheric velocity of $v_{\rm ph} = 29\,7000\pm1500$~km~s$^{-1}$.  This is more than $1\sigma$ times more rapid than the ``typical'' GRB-SN \citep{Cano2016}, which has $v_{\rm ph} = 20\,0000\pm1500$~km~s$^{-1}$, with a standard deviation of $\sigma=8000$~km~s$^{-1}$.  Such a rapid expansion velocity was also confirmed by \citet{Ashall17}.

To determine what powers the luminosity of SN~2016jca, we considered two models: a radioactive heating model and a magnetar model.  The latter model was individually fit to both the X-ray and the $r$-band luminosity LCs, and the best-fitting parameters for the spin period and magnetic field determined from both frequency regimes were inconsistent with each other.  Moreover, the model was unable to reproduce the shallow decay in both the X-ray and optical LCs seen at late times. The radioactive-heating model provided a much better fit to the $griz$ bolometric LC, where we found a nickel mass of $M_{\rm Ni} = 0.22\pm0.08$~M$_{\odot}$, and ejecta mass of $M_{\rm ej} = 5.8\pm0.3$~M$_{\odot}$, and a kinetic energy of $E_{\rm K} = 5.1\pm0.8\times10^{52}$~erg.   We also found a $\gamma$-ray opacity of $\kappa_{\gamma} = 0.034$~cm$^2$~g$^{-1}$, which falls within the range of $\gamma$-ray opacities found by \citet{Wheeler15} for a sample of $N=20$ SNe Ibc.

The kinetic energy found here is well in excess of that expected for an explosion powered by a magnetar, where a maximum value of $E_{\rm K} \leq 2\times10^{52}$~erg has been suggested \citep{Usov92,Mazzali14}.  The SN's energetics were also confirmed by \citet{Ashall17} to be more than $5\times10^{52}$~erg.  The results of our magnetar model, as well as the Fe \textsc{ii} $\lambda5169$ velocity evolution (which is not flat, as predicted by 1D magnetar models) also argue against a magnetar powering any phases of GRB~161219B/SN~2016jca.  Instead, the energetics found here are more indicative of a black hole being formed at the time of core-collapse.  This conclusion is at odds to that of \citet{Ashall17}, who despite this energetic constraint, argue that a magnetar may likely be powering the SN outflow, which is based at least on part on the large jet angle inferred from their observations.

When analysing the $\gamma$-ray properties of GRB~161219B, we found that its isotropic-equivalent $\gamma$-ray energy is $E_{\rm \gamma,iso} \approx 8.5\times10^{49}$~erg, which when we consider its duration ($T_{90}=6.9$~s), implies that it is an intermediate-luminosity GRB.  Moreover, along with its peak energy ($E_{\rm p} \approx 106$~keV), GRB~161219B is an outlier in the Amati relation.  We found the same conclusion when we considered the rest-frame energetics as constrained from both \emph{Swift}-BAT and Konus-\emph{Wind}.

The host galaxy of GRB~161219B/SN~2016jca appears to be an edge-on spiral, whose photometric ($grizY$) SED is consistent with a galaxy of age $\approx 1$~Gyr, a mass of $\approx 7.6 \times 10^8$~M$_{\odot}$, a SFR of $\approx 0.25$~M$_{\odot}$~yr$^{-1}$, and negligible intrinsic extinction.  Inspection of \emph{HST} images reveals that the GRB occurred at a projected distance of $\approx 3.4$~kpc from the host's center.  We divided the integrated host spectrum into discrete bins, and determined the metallicity, SFR and sSFR as a function of position along the galactic disk.  There is little variation in these values across the galaxy.  Neither the metallicity nor the SFR is extreme at the GRB's position compared with the rest of the host.  The SFR and sSFR are largest in the SF region at the opposite side of the galaxy.  From the integrated host spectrum we find a metallicity of $\approx 0.4$ solar, a modest SFR of $\approx 0.17$~M$_{\odot}$~yr$^{-1}$ and a SSFR of $\approx 2.91$~M$_{\odot}$~yr$^{-1}$~L~L$_*^{-1}$.   Both the mass and sSFR are commensurate with GRB hosts at $z<0.5$.  The derived host-integrated metallicity is perfectly commensurate with those of other GRB hosts. 

Finally, we report on the presence of a chromatic, pre-maximum bump in the observer-frame $g$-band filter.  At +1.5~days, an excess of flux in the $g$-band is seen in the XS spectrum.  The evolution of the $g$-band excess was shown by a time-sequence of UV to NIR SEDs, which appears to peak around $3-5$~days (observer-frame), and disappears by +7.6~days. While pre-maximum bumps have been seen for GRB-SNe, SLSNe-Ic and non-GRB SNe Ic, their achromatic behaviour means that the analytical models used to describe their physics (which are usually best-fit by the low-mass extended-envelope model of \citealt{Nakar15}) do not apply in this case.  After demonstrating that the $g$-band excess is real, we are unable to conclude on its physical origin.

In conclusion, SN~2016jca is only the seventh GRB-SNe to have been detected within 1~Gpc, and has therefore provided a rare but excellent chance to determine both its physical and observations properties.  In relation to the general GRB-SN population, its nickel mass and ejecta mass are perfectly commensurate.  However, its photospheric velocity appears to be more rapid than most GRB-SNe, which in turn implies a large kinetic energy.  Its large kinetic energy, taken in tandem with the results of the magnetar modelling and the velocity evolution, argue against a magnetar powering this event, and instead it is more likely that a black hole was formed at the time of core-collapse.

\section{Acknowledgements}

LI, CT, ZC, AdUP and DAK acknowledge support from the Spanish research project AYA 2014-58381-P. CT and AdUP furthermore acknowledge support from Ram\'on y Cajal fellowships RyC-2012-09984 and RyC-2012-09975. DAK and ZC acknowledge support from Juan de la Cierva Incorporaci\'on fellowships IJCI-2015-26153 and IJCI-2014-21669. RSR is supported by a 2016 BBVA Foundation Grant for Researchers and Cultural Creators. T.-W. Chen and T. Kr\"uhler acknowledge the support through the Sofia Kovalevskaja Award to P. Schady from the Alexander von Humboldt Foundation of Germany. DM acknowledges support from the Instrument center for Danish Astrophysics (IDA).

Our analysis was based on: (1) observations collected at the European Organisation for Astronomical Research in the Southern Hemisphere under ESO programme 098.A-0055(A). (2) Observations made with the Nordic Optical Telescope, operated by the Nordic Optical Telescope Scientific Association at the Observatorio del Roque de los Muchachos, La Palma, Spain, of the Instituto de Astrofisica de Canarias. (3) Observations collected at the European Organisation for Astronomical Research in the Southern Hemisphere, Chile as part of PESSTO, (the Public ESO Spectroscopic Survey for Transient Objects Survey) ESO program 188.D-3003, 191.D-0935, 197.D-1075. (4) Development of CIRCE was supported by the University of Florida and the National Science Foundation (grant AST-0352664), in collaboration with IUCAA. (5) Part of the funding for GROND (both hardware as well as personnel) was generously granted from the Leibniz-Prize to Prof. G. Hasinger (DFG grant HA 1850/28-1).  (6) The Pan-STARRS1 Surveys (PS1) and the PS1 public science archive have been made possible through contributions by the Institute for Astronomy, the University of Hawaii, the Pan-STARRS Project Office, the Max-Planck Society and its participating institutes, the Max Planck Institute for Astronomy, Heidelberg and the Max Planck Institute for Extraterrestrial Physics, Garching, The Johns Hopkins University, Durham University, the University of Edinburgh, the Queen's University Belfast, the Harvard-Smithsonian Center for Astrophysics, the Las Cumbres Observatory Global Telescope Network Incorporated, the National Central University of Taiwan, the Space Telescope Science Institute, the National Aeronautics and Space Administration under Grant No. NNX08AR22G issued through the Planetary Science Division of the NASA Science Mission Directorate, the National Science Foundation Grant No. AST-1238877, the University of Maryland, Eotvos Lorand University (ELTE), the Los Alamos National Laboratory, and the Gordon and Betty Moore Foundation.

%098.D-0710(C) Klaas's FORS

\begin{appendix}

\section{The Radioactive-Heating Model}
\label{sec:app_radioactive}

The radioactive-heating model used in this work is based on the original analytical model of \citet{Arnett1982}.  Since this seminal work, the basic model has been extended to include not only energy deposited via the radioactive decay of nickel, but also radioactive cobalt \citep{Valenti08}.  A further amendment to the model was made by \citet{Chatzopoulos11} to include a term that considers the leakage of $\gamma$-rays into space, and hence not depositing this energy into the expanding SN ejecta.

In the original \citet{Arnett1982} model there were several assumptions, many of which are still contained in the analytical model used here, which include:

\begin{enumerate}
\item A homologous expansion ($t^{-2}$ scaling) of the ejecta
\item Spherical symmetry
\item A photosphere that has a unique position in space
\item The radioactive material present in the ejecta is located at the centre of the explosion and does not mix
\item Radiation-pressure dominance
\item A small initial radius before explosion ($\rm{R_{0}} \rightarrow 0$)
\item The applicability of the diffusion approximation for photons (i.e. the \emph{photospheric phase})
\end{enumerate}

Caveats of these assumptions, and their effect on the resultant modelling results can be found in \citet{Cano13}.

The luminosity of a type I SNe as a function of time is:

\begin{footnotesize}
\begin{center}
\begin{equation}
\label{equ:bol1}
 L(t) = M_{\rm Ni}e^{-x^2}~{\left((\epsilon_{\rm Ni} - \epsilon_{\rm Co}) \int_{0}^{x}A(z)dz+ \epsilon_{\rm Co}\int_{0}^{x}B(z)dz\right)}~(1-e^{-Ct^{-2}})
\end{equation} 
\end{center}
\end{footnotesize}

\noindent where

\begin{center}
\begin{equation}
\label{equ:bol2}
A(z)=2ze^{-2zy+z^2}, B(z)=2ze^{-2zy+2zs+z^2} 
\end{equation} 
\end{center}

\noindent and $x\equiv t/\tau_{m}$, $y\equiv \tau_{m}/(2\tau_{Ni})$, and $s\equiv (\tau_{m}(\tau_{Co}-\tau_{Ni})/(2\tau_{Co}\tau_{Ni}))$.  

The factor ($1-e^{-Ct^{-2}}$) takes into consideration the possibility that some of the $\gamma$-rays produced during the radioactive decays escape directly into space, and hence do not interact with the SN ejecta.  Small values of $C$ imply that most of the $\gamma$-rays escape into space.  The $\gamma$-ray optical depth of the ejecta is $\tau = \kappa_{\gamma} \rho R = Ct^{-2}$, and hence the $\gamma$-ray opacity is $\kappa_{\gamma} = (4\pi C v_{\rm ph}^2) / (3 M_{\rm ej})$.

The energy release in one second by one gram of $^{56}$Ni and $^{56}$Co are, respectively, $\epsilon_{\rm Ni}=3.90 \times 10^{10}$ erg s$^{-1}$ g$^{-1}$ and $\epsilon_{\rm Co}=6.78 \times 10^{9}$ erg s$^{-1}$ g$^{-1}$ \citep{Sutherland84,Cappellaro97}.  The decay times of $^{56}$Ni and $^{56}$Co, respectively, are $\tau_{\rm Ni}=8.77$ days (see \citealt{Taubenberger06} and references therein) and  $\tau_{\rm Co}=111.3$ days \citep{Martin87}.  

$\tau_{m}$ is the effective diffusion time and determines the overall width of the bolometric light curve.  $\tau_{m}$ is expressed in relation to the  opacity $\kappa$ and the ejecta mass $\rm{M_{ej}}$, as well as the photospheric velocity ${v_{\rm ph}}$ at the time of bolometric maximum:

\begin{center}
\begin{equation}
\label{equ:tau} 
\tau_{m} \approx \left(\frac{\kappa}{\beta c}\right)^{1/2} \left(\frac{{M_{\rm ej}}}{{v_{\rm ph}}}\right)^{1/2}
\end{equation} 
\end{center}
%
%\noindent 
\noindent where $\beta \approx 13.8$ is a constant of integration \citep{Arnett1982}, and $c$ is the speed of light. Additionally, we assume a constant opacity $\kappa=0.07$ cm$^{2}$g$^{-1}$ (e.g. \citealt{Chugai00}), which is justified if electron scattering is the dominant opacity source (e.g. \citealt{Chevalier92}).  Finally, the kinetic energy of the ejecta is simply ${E_{\rm k}} = \frac{1}{2} M_{\rm ej} v_{\rm ph}^{2}$.

\section{The Magnetar Model}
\label{sec:app_magnetar}

The magnetar model used here is identical to that employed in \citet{CJM16}, in which the complete derivation of the model can be consulted.  For the sake of completeness, we represent the main features of the model here.

The model considers three phases: (1) An AG component arising from the initial collision of the GRB ejecta with the surrounding medium, (2) A magnetar-powered AG phase, and (3) a magnetar-powered SN phase.

Phase (1) is modelled as a SPL (e.g. \citealt{Rowlinson13,Cano15}), which is analogous to the impulsive energy input term in the model of \citet{ZhangMesz01}:

\begin{equation}
L_{\rm SPL} (t) = \Lambda t^{-\alpha} \hspace{5pt} ({\rm erg~s^{-1}})
\label{equ:SPL}
\end{equation}

\noindent where $\Lambda$ is the normalisation constant and $\alpha$ is the decay constant.  Here we assume $\alpha = \Gamma_{\gamma} + 1$, where $\Gamma_{\gamma}$ is the photon index of the prompt emission, assuming that the decay slope is governed by the curvature effect (e.g. \citealt{KumarPan00,Piran04}).

The magnetar-powered AG (which persists as long as the jet remains collimated enough to deposit energy into the expanding fireball at large radii, and not into the expanding SN) is modelled as a form of continuous energy input \citep{ZhangMesz01}.  The general idea here is a magnetar central engine that deposits Poynting flux dominated dipole radiation into the ejecta (e.g. \citealt{Dallosso11}) as:

\begin{equation}
L_{\rm AG}(t)=L_{0}\left(1+\frac{t}{T_{0}}\right)^{-2} \hspace{5pt} ({\rm erg~s^{-1}})
\label{equ:magnetar_AG}
\end{equation}

\noindent where $L_{0}$ is the plateau luminosity, $T_{0}$ is the plateau duration.  In order to reduce the amount of free-parameters we have assumed a canonical NS with a mass of $1.4$~M$_{\odot}$ and a radius of $10^{6}$~cm.

Once the jet spreads, it can no longer maintain a hole in the expanding ejecta, and instead it deposits its energy more locally in the SN itself.  The analytical prescription used here is based on the previous works of \citet{OstrikerGunn71}, \citet{KasenBild10}, \citet{Barkov11} and \citet{Chatzopoulos11}.  A magnetar-powered SN is expressed as:

\begin{equation}
 L_{\rm SN}(t) = \frac{E_{\rm p}}{t_{\rm p}}~{\rm exp}\left(\frac{-x^2}{2}\right)\int_0^x~\frac{z~{\rm exp}\left(\frac{z^2}{2}\right)}{(1+yz)^2}\, \mathrm{d}z \hspace{10pt} ({\rm erg~s^{-1}})
 \label{equ:mag_SN}
\end{equation}

\noindent where $E_{\rm p}$ is the initial energy of the magnetar (units of erg) and $t_{\rm p}$ is the characteristic spin-down time of the magnetar (units of days). Additionally, $x = t / t_{\rm diff}$ and $y = t_{\rm diff} / t_{\rm p}$, where $t_{\rm diff}$ is the diffusion timescale of the SN in units of days.  As in the magnetar-powered AG phase, the radius of the magnetar is assumed to be 10$^6$ cm (i.e. 10 km), and we considered an $l=2$ magnetic dipole.

From these models we can determine the initial spin-period ($P$) and magnetic-field strength ($B$) of the magnetar central engine:

\begin{equation}
B = \sqrt{\frac{1.3\times10^{2}~P^2}{t_{\rm p, yr}}} \hspace{5pt} ({\rm 10^{15}~G})
\label{equ:mag_SN_B}
\end{equation}

\noindent and

\begin{equation}
P = \sqrt{\frac{2\times10^{46}}{E_{\rm p}}} \hspace{5pt} ({\rm ms})
\label{equ:mag_SN_P}
\end{equation}

\noindent where $t_{\rm p, yr}$ is the characteristic spin-down time of the magnetar in units of years.

These three phases are combined into a single model:

\begin{equation}
 L_{\rm total} (t) = L_{\rm AG} + \Phi L_{\rm SN} + L_{\rm SPL}  \hspace{5pt} ({\rm erg~s^{-1}})
 \label{equ:mag_combined_norm_SN}
\end{equation}

\noindent where $\Phi$ is an additional free-parameter that was fit to the optical LCs.  Therefore, if a GRB-SN bump  has a value of $\Phi \approx 1$, this event can be considered as being powered entirely by EM emission from a magnetar central engine.  Conversely, for all events where $\Phi > 1$, additional sources of heating are needed to explain the luminosity of the SN phase, which is likely due to the heating from the radioactive decay of nickel and cobalt into their daughter products.

\section{Spectroscopic Observation Log}

A summary of our spectroscopic observations are given in Table \ref{table:spectra_obs_log}.

\begin{table*}
%\scriptsize
\centering
\setlength{\tabcolsep}{6.0pt}
\setlength{\extrarowheight}{3pt}
\caption{GRB 161219B / SN 2016jca: Spectroscopy observation log}
\label{table:spectra_obs_log}
\begin{tabular}{cccccc}
\hline													
UT date	&		$t-t_{0}$ (d)$^{a}$	&	Range (\AA)	&	Equipment	&	Exposure Time (s)	\\
\hline													
%24-Jun-2014	&	13:00:17	&	18.4089	&	-1.0	&	$3100-10300$	&	Keck, LRIS,  600/4000 (blue) grism \& 400/8500 (red) grating	&	1200 s	\\
21-Dec-2016 &  1.504 &  $3200-22000$ & VLT-XS & $4 \times 600$ \\
26-Dec-2016 &  7.245 &  $3700-7800$ & GTC-OSIRIS & $3 \times 900$ (in R1000B) \\
01-Jan-2017 &  13.253 &  $3700-9300$ & GTC-OSIRIS & $2 \times 600$ (in R1000B and R1000R each) \\
03-Jan-2018 &  15.455 &  $3985-9315$ & NTT-EFOSC2 & $2 \times 2700$ (grism 13) \\
09-Jan-2017 &  21.253 &  $3700-9300$ & GTC-OSIRIS & $2 \times 900$ (in R1000B and R1000R each) \\
16-Jan-2017 &  28.237 &  $3700-9300$ & GTC-OSIRIS & $2 \times 900$ (in R1000B and R1000R each) \\
22-Jan-2017 &  34.201 &  $3700-7800$ & GTC-OSIRIS & $4 \times 900$ (in R1000B) \\
26-Jan-2017 &  38.178 &  $3700-9300$ & GTC-OSIRIS & $2 \times 1200$ (in R1000B and R1000R each) \\
08-Feb-2017 &  51.144 &  $3700-9300$ & GTC-OSIRIS & $2 \times 1200$ (in R1000B and R1000R each) \\
28-Feb-2017 &  71.089 &  $3700-7800$ & GTC-OSIRIS & $2 \times 1200$ (in R1000B) \\
\hline	
\end{tabular}
%\medskip
\begin{flushleft}
$^{a}$ UT start time. \\
%$^{b}$ Relative to peak, observer-frame $r$-band light. \\
\end{flushleft}
\end{table*}

\section{Photometry}

Our optical/NIR photometry is presented in Table \ref{table:photometry}.  All magnitudes are of the AG+SN+host galaxy, are uncorrected for extinction, and are for a $2\farcs2$ circular aperture centered on the position of the OT.  Magnitudes in filters $griz$ are in the AB system, while those in filters $JHK$ are in Vega.

%\cleardoublepage

%\input{Photometry_Table.tex}

\newpage

\begin{landscape}
\begin{table}
\scriptsize
\centering
\setlength{\tabcolsep}{4.0pt}
\setlength{\extrarowheight}{0pt}
\caption{GRB 161219B/SN 2016jca - Photometry}
\label{table:photometry}
\begin{tabular}{ccccc|ccccc|ccccc|ccccc}
Telescope	&	date	&	filter	&	$t-t_0$ (d)	&		mag				&	Telescope	&	date	&	filter	&	$t-t_0$(d)	&		mag				&	Telescope	&	date	&	filter	&	$t-t_0$ (d)	&		mag				&	Telescope	&	date	&	filter	&	$t-t_0$ (d)	&		mag				\\
\hline																																																							
GROND	&	20-Dec	&	$g$	&	0.285	&	$	18.02	\pm	0.02	$	&	GROND	&	30-Dec	&	$r$	&	10.306	&	$	19.63	\pm	0.02	$	&	GTC	&	09-Jan	&	$i$	&	21.282	&	$	19.72	\pm	0.03	$	&	GROND	&	18-Feb	&	$z$	&	60.340	&	$	20.58	\pm	0.03	$	\\
GROND	&	20-Dec	&	$g$	&	0.300	&	$	18.15	\pm	0.02	$	&	GROND	&	31-Dec	&	$r$	&	11.338	&	$	19.65	\pm	0.02	$	&	GROND	&	11-Jan	&	$i$	&	22.286	&	$	19.70	\pm	0.03	$	&	NOT	&	27-Feb	&	$z$	&	70.115	&	$	20.56	\pm	0.06	$	\\
GROND	&	20-Dec	&	$g$	&	0.311	&	$	18.16	\pm	0.02	$	&	NOT	&	31-Dec	&	$r$	&	12.259	&	$	19.64	\pm	0.05	$	&	GROND	&	14-Jan	&	$i$	&	25.378	&	$	19.82	\pm	0.05	$	&	GTC	&	28-Feb	&	$z$	&	71.147	&	$	20.51	\pm	0.05	$	\\
GROND	&	21-Dec	&	$g$	&	1.557	&	$	19.27	\pm	0.03	$	&	GROND	&	01-Jan	&	$r$	&	12.291	&	$	19.65	\pm	0.02	$	&	NOT	&	16-Jan	&	$i$	&	28.167	&	$	19.90	\pm	0.02	$	&	PS1	&	-	&	$z$	&	-	&	$	20.70	\pm	0.09	$	\\
GROND	&	22-Dec	&	$g$	&	2.561	&	$	19.43	\pm	0.02	$	&	NOT	&	01-Jan	&	$r$	&	13.213	&	$	19.67	\pm	0.02	$	&	NOT	&	18-Jan	&	$i$	&	30.204	&	$	19.92	\pm	0.03	$	&	GROND	&	20-Dec	&	$J$	&	0.283	&	$	16.77	\pm	0.03	$	\\
GROND	&	24-Dec	&	$g$	&	4.504	&	$	19.52	\pm	0.02	$	&	GTC	&	01-Jan	&	$r$	&	13.278	&	$	19.74	\pm	0.03	$	&	GROND	&	20-Jan	&	$i$	&	31.500	&	$	20.03	\pm	0.02	$	&	GROND	&	20-Dec	&	$J$	&	0.298	&	$	16.82	\pm	0.03	$	\\
GROND	&	25-Dec	&	$g$	&	5.501	&	$	19.59	\pm	0.03	$	&	GROND	&	02-Jan	&	$r$	&	13.345	&	$	19.69	\pm	0.02	$	&	NOT	&	22-Jan	&	$i$	&	34.166	&	$	20.13	\pm	0.02	$	&	GROND	&	20-Dec	&	$J$	&	0.308	&	$	16.90	\pm	0.03	$	\\
GROND	&	26-Dec	&	$g$	&	6.488	&	$	19.76	\pm	0.03	$	&	GROND	&	03-Jan	&	$r$	&	14.358	&	$	19.69	\pm	0.02	$	&	GTC	&	22-Jan	&	$i$	&	34.232	&	$	20.09	\pm	0.03	$	&	GROND	&	21-Dec	&	$J$	&	1.554	&	$	17.86	\pm	0.03	$	\\
GTC	&	26-Dec	&	$g$	&	7.263	&	$	19.88	\pm	0.03	$	&	NOT	&	03-Jan	&	$r$	&	15.299	&	$	19.66	\pm	0.03	$	&	NOT	&	23-Jan	&	$i$	&	35.196	&	$	20.15	\pm	0.03	$	&	GROND	&	22-Dec	&	$J$	&	2.559	&	$	18.24	\pm	0.05	$	\\
NOT	&	27-Dec	&	$g$	&	8.245	&	$	19.89	\pm	0.02	$	&	GROND	&	04-Jan	&	$r$	&	15.370	&	$	19.70	\pm	0.02	$	&	NOT	&	25-Jan	&	$i$	&	37.189	&	$	20.18	\pm	0.03	$	&	GROND	&	24-Dec	&	$J$	&	4.502	&	$	18.71	\pm	0.06	$	\\
NOT	&	28-Dec	&	$g$	&	9.251	&	$	19.85	\pm	0.06	$	&	GROND	&	07-Jan	&	$r$	&	18.385	&	$	19.75	\pm	0.02	$	&	GROND	&	26-Jan	&	$i$	&	37.368	&	$	20.21	\pm	0.03	$	&	GROND	&	25-Dec	&	$J$	&	5.498	&	$	18.73	\pm	0.05	$	\\
GROND	&	29-Dec	&	$g$	&	9.296	&	$	19.86	\pm	0.03	$	&	GROND	&	09-Jan	&	$r$	&	20.336	&	$	19.83	\pm	0.02	$	&	GTC	&	26-Jan	&	$i$	&	38.215	&	$	20.15	\pm	0.02	$	&	GROND	&	26-Dec	&	$J$	&	6.487	&	$	18.57	\pm	0.09	$	\\
GROND	&	30-Dec	&	$g$	&	10.306	&	$	19.88	\pm	0.02	$	&	GTC	&	09-Jan	&	$r$	&	21.280	&	$	19.82	\pm	0.03	$	&	NOT	&	28-Jan	&	$i$	&	40.165	&	$	20.19	\pm	0.02	$	&	GROND	&	29-Dec	&	$J$	&	9.296	&	$	18.96	\pm	0.09	$	\\
GROND	&	31-Dec	&	$g$	&	11.338	&	$	19.96	\pm	0.04	$	&	GROND	&	11-Jan	&	$r$	&	22.286	&	$	19.95	\pm	0.02	$	&	GROND	&	31-Jan	&	$i$	&	42.345	&	$	20.25	\pm	0.03	$	&	GROND	&	30-Dec	&	$J$	&	10.306	&	$	18.86	\pm	0.07	$	\\
NOT	&	31-Dec	&	$g$	&	12.253	&	$	20.04	\pm	0.04	$	&	GROND	&	14-Jan	&	$r$	&	25.378	&	$	20.08	\pm	0.02	$	&	GROND	&	05-Feb	&	$i$	&	47.389	&	$	20.33	\pm	0.03	$	&	GROND	&	01-Jan	&	$J$	&	12.291	&	$	18.69	\pm	0.25	$	\\
GROND	&	01-Jan	&	$g$	&	12.291	&	$	19.92	\pm	0.03	$	&	NOT	&	16-Jan	&	$r$	&	28.159	&	$	20.15	\pm	0.02	$	&	GTC	&	08-Feb	&	$i$	&	51.186	&	$	20.49	\pm	0.03	$	&	GROND	&	02-Jan	&	$J$	&	13.345	&	$	18.80	\pm	0.26	$	\\
NOT	&	01-Jan	&	$g$	&	13.208	&	$	20.12	\pm	0.02	$	&	NOT	&	18-Jan	&	$r$	&	30.196	&	$	20.18	\pm	0.02	$	&	GROND	&	18-Feb	&	$i$	&	60.340	&	$	20.49	\pm	0.02	$	&	GROND	&	03-Jan	&	$J$	&	14.358	&	$	18.82	\pm	0.29	$	\\
GTC	&	01-Jan	&	$g$	&	13.276	&	$	20.12	\pm	0.03	$	&	GROND	&	20-Jan	&	$r$	&	31.500	&	$	20.27	\pm	0.02	$	&	NOT	&	27-Feb	&	$i$	&	70.101	&	$	20.55	\pm	0.03	$	&	GROND	&	07-Jan	&	$J$	&	18.385	&	$	18.91	\pm	0.09	$	\\
GROND	&	02-Jan	&	$g$	&	13.345	&	$	19.99	\pm	0.03	$	&	NOT	&	22-Jan	&	$r$	&	34.156	&	$	20.35	\pm	0.02	$	&	GTC	&	28-Feb	&	$i$	&	71.145	&	$	20.45	\pm	0.03	$	&	GROND	&	09-Jan	&	$J$	&	20.336	&	$	19.06	\pm	0.10	$	\\
GROND	&	03-Jan	&	$g$	&	14.358	&	$	20.24	\pm	0.03	$	&	GTC	&	22-Jan	&	$r$	&	34.230	&	$	20.33	\pm	0.02	$	&	PS1	&	-	&	$i$	&	-	&	$	21.08	\pm	0.06	$	&	NOT	&	09-Jan	&	$J$	&	21.220	&	$	18.75	\pm	0.18	$	\\
NOT	&	03-Jan	&	$g$	&	15.291	&	$	20.22	\pm	0.02	$	&	NOT	&	23-Jan	&	$r$	&	35.210	&	$	20.39	\pm	0.02	$	&	GROND	&	20-Dec	&	$z$	&	0.285	&	$	17.86	\pm	0.03	$	&	GROND	&	11-Jan	&	$J$	&	22.285	&	$	18.76	\pm	0.28	$	\\
GROND	&	04-Jan	&	$g$	&	15.370	&	$	20.20	\pm	0.03	$	&	NOT	&	25-Jan	&	$r$	&	37.181	&	$	20.44	\pm	0.02	$	&	GROND	&	20-Dec	&	$z$	&	0.300	&	$	17.90	\pm	0.03	$	&	GROND	&	14-Jan	&	$J$	&	25.375	&	$	19.03	\pm	0.08	$	\\
GROND	&	07-Jan	&	$g$	&	18.385	&	$	20.55	\pm	0.03	$	&	GROND	&	26-Jan	&	$r$	&	37.368	&	$	20.43	\pm	0.03	$	&	GROND	&	20-Dec	&	$z$	&	0.311	&	$	17.90	\pm	0.02	$	&	GROND	&	20-Jan	&	$J$	&	31.498	&	$	18.99	\pm	0.09	$	\\
GROND	&	09-Jan	&	$g$	&	20.336	&	$	20.67	\pm	0.04	$	&	GTC	&	26-Jan	&	$r$	&	38.213	&	$	20.50	\pm	0.04	$	&	GROND	&	21-Dec	&	$z$	&	1.557	&	$	19.01	\pm	0.03	$	&	GROND	&	26-Jan	&	$J$	&	37.365	&	$	19.02	\pm	0.18	$	\\
GTC	&	09-Jan	&	$g$	&	21.278	&	$	20.58	\pm	0.04	$	&	NOT	&	28-Jan	&	$r$	&	40.157	&	$	20.46	\pm	0.02	$	&	GROND	&	22-Dec	&	$z$	&	2.561	&	$	19.23	\pm	0.03	$	&	GROND	&	31-Jan	&	$J$	&	42.342	&	$	19.19	\pm	0.09	$	\\
GROND	&	11-Jan	&	$g$	&	22.286	&	$	20.77	\pm	0.05	$	&	GROND	&	31-Jan	&	$r$	&	42.345	&	$	20.63	\pm	0.02	$	&	GROND	&	24-Dec	&	$z$	&	4.505	&	$	19.51	\pm	0.02	$	&	GROND	&	05-Feb	&	$J$	&	47.386	&	$	19.46	\pm	0.10	$	\\
GROND	&	14-Jan	&	$g$	&	25.378	&	$	20.90	\pm	0.05	$	&	GROND	&	05-Feb	&	$r$	&	47.389	&	$	20.61	\pm	0.02	$	&	GROND	&	25-Dec	&	$z$	&	5.501	&	$	19.59	\pm	0.02	$	&	GTC	&	08-Feb	&	$J$	&	51.202	&	$	19.51	\pm	0.07	$	\\
NOT	&	16-Jan	&	$g$	&	28.151	&	$	20.88	\pm	0.02	$	&	GTC	&	08-Feb	&	$r$	&	51.184	&	$	20.73	\pm	0.05	$	&	GROND	&	26-Dec	&	$z$	&	6.488	&	$	19.65	\pm	0.04	$	&	GROND	&	18-Feb	&	$J$	&	60.337	&	$	19.40	\pm	0.27	$	\\
NOT	&	18-Jan	&	$g$	&	30.188	&	$	20.96	\pm	0.03	$	&	GROND	&	18-Feb	&	$r$	&	60.340	&	$	20.78	\pm	0.02	$	&	GTC	&	26-Dec	&	$z$	&	7.276	&	$	19.69	\pm	0.02	$	&	GROND	&	20-Dec	&	$H$	&	0.282	&	$	16.13	\pm	0.03	$	\\
GROND	&	20-Jan	&	$g$	&	31.500	&	$	21.22	\pm	0.03	$	&	NOT	&	27-Feb	&	$r$	&	70.090	&	$	20.83	\pm	0.02	$	&	NOT	&	27-Dec	&	$z$	&	8.271	&	$	19.70	\pm	0.04	$	&	GROND	&	20-Dec	&	$H$	&	0.298	&	$	16.20	\pm	0.04	$	\\
NOT	&	22-Jan	&	$g$	&	34.145	&	$	21.02	\pm	0.03	$	&	GTC	&	28-Feb	&	$r$	&	71.143	&	$	20.84	\pm	0.04	$	&	NOT	&	28-Dec	&	$z$	&	9.280	&	$	19.70	\pm	0.09	$	&	GROND	&	20-Dec	&	$H$	&	0.309	&	$	16.14	\pm	0.03	$	\\
GTC	&	22-Jan	&	$g$	&	34.229	&	$	21.05	\pm	0.03	$	&	PS1	&	-	&	$r$	&	-	&	$	21.08	\pm	0.05	$	&	GROND	&	29-Dec	&	$z$	&	9.296	&	$	19.61	\pm	0.03	$	&	GROND	&	21-Dec	&	$H$	&	1.554	&	$	17.15	\pm	0.04	$	\\
NOT	&	23-Jan	&	$g$	&	35.216	&	$	21.10	\pm	0.02	$	&	GROND	&	20-Dec	&	$i$	&	0.285	&	$	17.94	\pm	0.03	$	&	GROND	&	30-Dec	&	$z$	&	10.306	&	$	19.62	\pm	0.04	$	&	GROND	&	22-Dec	&	$H$	&	2.559	&	$	17.58	\pm	0.06	$	\\
NOT	&	25-Jan	&	$g$	&	37.173	&	$	21.05	\pm	0.03	$	&	GROND	&	20-Dec	&	$i$	&	0.300	&	$	17.97	\pm	0.02	$	&	GROND	&	31-Dec	&	$z$	&	11.338	&	$	19.70	\pm	0.03	$	&	GROND	&	24-Dec	&	$H$	&	4.502	&	$	18.07	\pm	0.07	$	\\
GROND	&	26-Jan	&	$g$	&	37.368	&	$	21.24	\pm	0.03	$	&	GROND	&	20-Dec	&	$i$	&	0.311	&	$	17.99	\pm	0.01	$	&	GROND	&	01-Jan	&	$z$	&	12.291	&	$	19.73	\pm	0.03	$	&	GROND	&	25-Dec	&	$H$	&	5.498	&	$	18.18	\pm	0.08	$	\\
GTC	&	26-Jan	&	$g$	&	38.209	&	$	21.13	\pm	0.02	$	&	GROND	&	21-Dec	&	$i$	&	1.557	&	$	19.07	\pm	0.02	$	&	GTC	&	01-Jan	&	$z$	&	13.281	&	$	19.76	\pm	0.03	$	&	GROND	&	26-Dec	&	$H$	&	6.487	&	$	18.06	\pm	0.12	$	\\
NOT	&	28-Jan	&	$g$	&	40.149	&	$	21.16	\pm	0.03	$	&	GROND	&	22-Dec	&	$i$	&	2.561	&	$	19.37	\pm	0.02	$	&	GROND	&	02-Jan	&	$z$	&	13.345	&	$	19.80	\pm	0.03	$	&	GROND	&	29-Dec	&	$H$	&	9.296	&	$	18.49	\pm	0.14	$	\\
GROND	&	31-Jan	&	$g$	&	42.345	&	$	21.22	\pm	0.04	$	&	GROND	&	24-Dec	&	$i$	&	4.504	&	$	19.69	\pm	0.02	$	&	GROND	&	03-Jan	&	$z$	&	14.358	&	$	19.77	\pm	0.03	$	&	GROND	&	01-Jan	&	$H$	&	12.291	&	$	18.40	\pm	0.11	$	\\
GROND	&	05-Feb	&	$g$	&	47.389	&	$	21.45	\pm	0.04	$	&	GROND	&	25-Dec	&	$i$	&	5.501	&	$	19.76	\pm	0.02	$	&	NOT	&	03-Jan	&	$z$	&	15.325	&	$	19.76	\pm	0.03	$	&	GROND	&	02-Jan	&	$H$	&	13.345	&	$	18.27	\pm	0.10	$	\\
GTC	&	08-Feb	&	$g$	&	51.183	&	$	21.23	\pm	0.07	$	&	GROND	&	26-Dec	&	$i$	&	6.488	&	$	19.74	\pm	0.03	$	&	GROND	&	04-Jan	&	$z$	&	15.370	&	$	19.79	\pm	0.04	$	&	GROND	&	03-Jan	&	$H$	&	14.358	&	$	18.34	\pm	0.14	$	\\
GROND	&	18-Feb	&	$g$	&	60.340	&	$	21.39	\pm	0.04	$	&	GTC	&	26-Dec	&	$i$	&	7.273	&	$	19.81	\pm	0.02	$	&	GROND	&	07-Jan	&	$z$	&	18.385	&	$	19.77	\pm	0.03	$	&	GROND	&	09-Jan	&	$H$	&	20.336	&	$	18.32	\pm	0.09	$	\\
NOT	&	27-Feb	&	$g$	&	70.078	&	$	21.45	\pm	0.03	$	&	NOT	&	27-Dec	&	$i$	&	8.261	&	$	19.75	\pm	0.02	$	&	GROND	&	09-Jan	&	$z$	&	20.336	&	$	19.75	\pm	0.03	$	&	NOT	&	09-Jan	&	$H$	&	21.245	&	$	18.45	\pm	0.16	$	\\
GTC	&	28-Feb	&	$g$	&	71.135	&	$	21.26	\pm	0.04	$	&	NOT	&	28-Dec	&	$i$	&	9.270	&	$	19.59	\pm	0.03	$	&	GTC	&	09-Jan	&	$z$	&	21.284	&	$	19.93	\pm	0.04	$	&	GROND	&	11-Jan	&	$H$	&	22.285	&	$	18.12	\pm	0.09	$	\\
PS1	&	-	&	$g$	&	-	&	$	21.69	\pm	0.07	$	&	GROND	&	29-Dec	&	$i$	&	9.296	&	$	19.69	\pm	0.02	$	&	GROND	&	11-Jan	&	$z$	&	22.286	&	$	19.79	\pm	0.04	$	&	GROND	&	14-Jan	&	$H$	&	25.375	&	$	18.74	\pm	0.12	$	\\
GROND	&	20-Dec	&	$r$	&	0.286	&	$	18.07	\pm	0.01	$	&	GROND	&	30-Dec	&	$i$	&	10.306	&	$	19.65	\pm	0.02	$	&	GROND	&	14-Jan	&	$z$	&	25.378	&	$	19.87	\pm	0.03	$	&	GROND	&	20-Jan	&	$H$	&	31.498	&	$	18.78	\pm	0.13	$	\\
GROND	&	20-Dec	&	$r$	&	0.300	&	$	18.10	\pm	0.01	$	&	GROND	&	31-Dec	&	$i$	&	11.338	&	$	19.64	\pm	0.03	$	&	NOT	&	16-Jan	&	$z$	&	28.177	&	$	20.03	\pm	0.04	$	&	GROND	&	31-Jan	&	$H$	&	42.342	&	$	18.83	\pm	0.16	$	\\
GROND	&	20-Dec	&	$r$	&	0.311	&	$	18.12	\pm	0.01	$	&	NOT	&	31-Dec	&	$i$	&	12.264	&	$	19.56	\pm	0.12	$	&	NOT	&	18-Jan	&	$z$	&	30.214	&	$	20.02	\pm	0.04	$	&	GROND	&	05-Feb	&	$H$	&	47.386	&	$	18.91	\pm	0.13	$	\\
GROND	&	21-Dec	&	$r$	&	1.556	&	$	19.25	\pm	0.02	$	&	GROND	&	01-Jan	&	$i$	&	12.291	&	$	19.67	\pm	0.02	$	&	GROND	&	20-Jan	&	$z$	&	31.500	&	$	20.15	\pm	0.03	$	&	GTC	&	08-Feb	&	$H$	&	51.206	&	$	19.11	\pm	0.05	$	\\
GROND	&	22-Dec	&	$r$	&	2.561	&	$	19.57	\pm	0.01	$	&	NOT	&	01-Jan	&	$i$	&	13.217	&	$	19.68	\pm	0.03	$	&	NOT	&	22-Jan	&	$z$	&	34.179	&	$	20.11	\pm	0.04	$	&	GROND	&	18-Feb	&	$H$	&	60.337	&	$	18.59	\pm	0.25	$	\\
GROND	&	24-Dec	&	$r$	&	4.505	&	$	19.75	\pm	0.02	$	&	GTC	&	01-Jan	&	$i$	&	13.279	&	$	19.70	\pm	0.03	$	&	GTC	&	22-Jan	&	$z$	&	34.234	&	$	20.12	\pm	0.03	$	&	GROND	&	20-Dec	&	$K$	&	0.283	&	$	15.84	\pm	0.05	$	\\
GROND	&	25-Dec	&	$r$	&	5.501	&	$	19.74	\pm	0.02	$	&	GROND	&	02-Jan	&	$i$	&	13.345	&	$	19.63	\pm	0.02	$	&	NOT	&	25-Jan	&	$z$	&	37.199	&	$	20.19	\pm	0.04	$	&	GROND	&	20-Dec	&	$K$	&	0.296	&	$	15.70	\pm	0.04	$	\\
GROND	&	26-Dec	&	$r$	&	6.487	&	$	19.71	\pm	0.03	$	&	GROND	&	03-Jan	&	$i$	&	14.358	&	$	19.61	\pm	0.02	$	&	GROND	&	26-Jan	&	$z$	&	37.368	&	$	20.19	\pm	0.04	$	&	GROND	&	20-Dec	&	$K$	&	0.308	&	$	15.72	\pm	0.05	$	\\
GTC	&	26-Dec	&	$r$	&	7.270	&	$	19.78	\pm	0.03	$	&	NOT	&	03-Jan	&	$i$	&	15.307	&	$	19.71	\pm	0.02	$	&	GTC	&	26-Jan	&	$z$	&	38.217	&	$	20.16	\pm	0.03	$	&	GROND	&	21-Dec	&	$K$	&	1.554	&	$	16.60	\pm	0.09	$	\\
NOT	&	27-Dec	&	$r$	&	8.253	&	$	19.69	\pm	0.02	$	&	GROND	&	04-Jan	&	$i$	&	15.370	&	$	19.65	\pm	0.02	$	&	NOT	&	28-Jan	&	$z$	&	40.175	&	$	20.41	\pm	0.08	$	&	GROND	&	22-Dec	&	$K$	&	2.557	&	$	17.09	\pm	0.12	$	\\
NOT	&	28-Dec	&	$r$	&	9.259	&	$	19.64	\pm	0.03	$	&	GROND	&	07-Jan	&	$i$	&	18.385	&	$	19.63	\pm	0.02	$	&	GROND	&	05-Feb	&	$z$	&	47.389	&	$	20.29	\pm	0.03	$	&	GROND	&	25-Dec	&	$K$	&	5.498	&	$	17.83	\pm	0.06	$	\\
GROND	&	29-Dec	&	$r$	&	9.296	&	$	19.68	\pm	0.02	$	&	GROND	&	09-Jan	&	$i$	&	20.336	&	$	19.63	\pm	0.02	$	&	GTC	&	08-Feb	&	$z$	&	51.188	&	$	20.42	\pm	0.04	$	&	GTC	&	08-Feb	&	$K$	&	51.210	&	$	18.66	\pm	0.05	$	\\
\hline													
\end{tabular}
%\medskip
\begin{flushleft}
NB: Photometry in $griz$ is AB, while $JHK$ are Vega.  All photometry is of the AG+SN+host, and are uncorrected for extinction.\\
$t_0 = 2457742.284$ in Julian date.
\end{flushleft}
\end{table}

\end{landscape}

\end{appendix}

%\label{lastpage}

\end{document}